\RequirePackage{rotating}
\documentclass{aa}
\usepackage{natbib}
\usepackage{amsmath}
\usepackage{gensymb}
\usepackage{verbatim}
\bibpunct{(}{)}{;}{a}{}{,} 
\newcommand{\gae}{\lower 2pt \hbox{$\, \buildrel {\scriptstyle >}\over {\scriptstyle \sim}\,$}}
\usepackage{graphicx}
\usepackage{caption}
\usepackage{multirow}
\usepackage{pgf}
\usepackage{float}
\usepackage[varg]{txfonts}
\usepackage{subfig}
\begin{document}

\title{Neon, sulphur and argon abundances of planetary nebulae in the sub-solar metallicity Galactic anti-centre}

\author{G. J. S. Pagomenos\inst{\ref{inst1}}, J. Bernard-Salas\inst{\ref{inst1}}, S. R. Pottasch\inst{\ref{inst2}}}

\institute{School of Physical Sciences, The Open University, Milton Keynes, MK7 6AA, UK \\ \email{george.pagomenos@open.ac.uk}\label{inst1}
\and Kapteyn Astronomical Institute, PO Box 800, 9700 AV Groningen, The Netherlands\label{inst2}}

\abstract
{Spectra of planetary nebulae show numerous fine structure emission lines from ionic species, enabling us to study the overall abundances of the nebular material that is ejected into the interstellar medium. The abundances derived from planetary nebula emission show the presence of a metallicity gradient within the disk of the Milky Way up to Galactocentric distances of $\sim$~10~kpc, which are consistent with findings from studies of different types of sources, including H~{\tiny II} regions and young B-type stars. The radial dependence of these abundances further from the Galactic centre is in dispute.}
{We aim to derive the abundances of neon, sulphur and argon from a sample of planetary nebulae towards the Galactic anti-centre, which represent the abundances of the clouds from which they were formed, as they remain unchanged throughout the course of stellar evolution.  We then aim to compare these values with similarly analysed data from elsewhere in the Milky Way in order to observe whether the abundance gradient continues in the outskirts of our Galaxy.}
{We have observed 23 planetary nebulae at Galactocentric distances of 8--21~kpc with \emph{Spitzer} IRS. The abundances were calculated from infrared emission lines, for which we observed the main ionisation states of neon, sulphur, and argon, which are little affected by extinction and uncertainties in temperature measurements or fluctuations within the planetary nebula. We have complemented these observations with others from optical studies in the literature, in order to reduce or avoid the need for ionisation correction factors in abundance calculations.}
{The overall abundances of our sample of planetary nebulae in the Galactic anti-centre are lower than those in the solar neighbourhood. The abundances of neon, sulphur, and argon from these stars are consistent with a metallicity gradient from the solar neighbourhood up to Galactocentric distances of $\sim$~20~kpc, albeit with varying degrees of dispersion within the data. }
{}

\keywords{stars: late-type -- stars: abundances, planetary nebulae: general, Galaxy: abundances, infrared: ISM -- infrared: stars}

\date{Received <date> / Accepted <date>}

\titlerunning{Ne, S, \& Ar abundances in Galactic anti-centre PNe}
\authorrunning{Pagomenos et al.}

\maketitle

\section{Introduction}

\begin{table*}[tp]
\caption{List of 23 PNe observed in the sample.}
\label{tab:23PNe}
\centering
\begin{tabular}{ c c c c c c c }
\hline\hline
Source & Source & RA (J2000) & Dec (J2000) & AORkey & AORKey & $R_{g}$ \\
\emph{Name} & \emph{PNG} & \emph{(h m s)} & \emph{(d m s)} & & \emph{off position} & \emph{(kpc)} \\
\hline 
J320 & $190.3-17.7$ & 05 05 34.32 & $+10$ 42 23.8 & 21946880 & 21947136 & 13.6 $\pm$ 1.6 \\
K3-65 & $153.7-01.4$ & 04 15 54.53 & $+48$ 49 40.1 & 21947392 & 21947648 & 11.5 $\pm$ 2.1* \\
K3-66 & $167.4-09.1$ & 04 36 37.23 & $+33$ 39 30.0 & 21947904 & 21948160 & 15.8 $\pm$ 2.3 \\
K3-67 & $165.5-06.5$ & 04 39 47.93 & $+36$ 45 42.6 & 21948416 & 21948672 & 14.4 $\pm$ 4.3$^{\dagger}$ \\
K3-68 & $178.3-02.5$ & 05 31 35.86 & $+28$ 58 41.6 & 21948928 & 21949184 & 10.2 $\pm$ 1.8* \\
K3-69 & $170.7+04.6$ & 05 41 22.13 & $+39$ 15 08.1 & 21949440 & 21949696 & $>$13.9* \\
K3-70 & $184.6+00.6$ & 05 58 45.34 & $+25$ 18 43.8 & 21949952 & 21950208 & $>$14.0* \\
K3-71 & $184.8+04.4$ & 06 13 54.98 & $+26$ 52 57.0 & 21950464 & 21950720 & 10.5 $\pm$ 2.0* \\
K3-90 & $126.3+02.9$ & 01 24 58.70 & $+65$ 38 34.7 & 21950976 & 21951232 & $<$8.7* \\
K4-48 & $201.7+02.5$ & 06 39 55.84 & $+11$ 06 30.3 & 21952000 & 21952256 & 16.6 $\pm$ 5.0$^{\dagger}$ \\
M1-1 & $130.3-11.7$ & 01 37 19.43 & $+50$ 28 11.6 & 21952512 & 21952768 & 14.7 $\pm$ 2.4 \\
M1-6 & $211.2-03.5$ & 06 35 45.13 & $-00$ 05 37.5 & 21953024 & 21953280 & 9.8 $\pm$ 1.8* \\
M1-7 & $189.8+07.7$ & 06 37 20.96 & $+24$ 00 35.4 & 21953536 & 21953792 & 14.5 $\pm$ 1.9 \\
M1-8 & $210.3+01.9$ & 06 53 33.79 & $+03$ 08 27.0 & 21954048 & 21954304 & 12.2 $\pm$ 1.4 \\
M1-9 & $212.0+04.3$ & 07 05 19.20 & $+02$ 46 59.5 & 21954560 & 21954816 & 16.2 $\pm$ 2.5 \\
M1-14 & $234.9-01.4$ & 07 27 56.50 & $-20$ 13 22.8 & 21955072 & 21955328 & 11.4 $\pm$ 1.1 \\
M1-16 & $226.7+05.6$ & 07 37 18.93 & $-09$ 38 48.0 & 21955584 & 21955840 & 13.0 $\pm$ 1.8 \\
M1-17 & $228.8+05.3$ & 07 40 22.19 & $-11$ 32 29.9 & 21956096 & 21956352 & 14.8 $\pm$ 2.3 \\
M2-2 & $147.8+04.1$ & 04 13 15.04 & $+56$ 56 58.1 & 21956608 & 21956864 & $>$9.7* \\
M3-2 & $240.3-07.6$ & 07 14 49.92 & $-27$ 50 23.3 & 21957120 & 21957376 & 15.3 $\pm$ 2.6 \\
M4-18 & $146.7+07.6$ & 04 25 50.85 & $+60$ 07 12.8 & 21957632 & 21957888 & 15.0 $\pm$ 2.2 \\
SaSt2-3 & $232.0+05.7$ & 07 48 03.67 & $-14$ 07 40.4 & 21958144 & 21958400 & 20.8 $\pm$ 4.1 \\
Y-C 2-5 & $240.3+07.0$ & 08 10 41.64 & $-20$ 31 32.6 & 21958656 & 21958912 & 13.2 $\pm$ 6.6$^{\ddagger}$ \\
\hline
\end{tabular}
\tablefoot{Galactocentric distances and their errors were determined from heliocentric values determined statistically from~\cite{frew16}, assuming $R_{g,\odot} =$ 8.0~$\pm$~0.5~kpc, except: * directly measured distances from~\cite{giammanco11}; $^{\dagger}$ statistical distances from~\cite{phillips11}; $^{\ddagger}$ statistical distance from~\cite{costa04}, from which we assume a 50\% error.}
\end{table*}

As Sun-like stars (of $\sim$~0.8--8~$M_{\odot}$) evolve, they eventually become planetary nebulae (PNe). In this evolutionary phase, the star has lost enough of its convective envelope through stellar winds to expose its inner, hotter regions, causing this ejected material to become ionised. We can determine the ionic and elemental abundances of this PN ejecta by analysing the strong forbidden emission lines of the ions in the stellar spectrum.

Much of the observable PN emission comes from ionisation states of $\alpha$-process elements, such as neon, sulphur, and argon. The main stages of ionisation for these three elements are observed at mid-infrared wavelengths. While the abundances of elements such as helium and carbon change significantly throughout the course of stellar evolution, those of neon, sulphur, and argon are unchanged (e.g.~\citealt{marigo03}), making them useful probes of metallicity at the epoch of stellar formation. Abundance studies carried out at optical wavelengths commonly use oxygen as a metric for metallicity, as the observed emission lines of O$^+$ and O$^{2+}$ are always strong. However, during the evolution of these low- to intermediate-mass stars, the abundances of oxygen are known to change. This particularly occurs within the asymptotic giant branch (AGB) phase, in which the third dredge-up brings helium, carbon and a small amount of oxygen to the outer envelope of the star. For stars with M~$\gtrsim$~4~M$_{\odot}$, some oxygen will also be destroyed by hot bottom burning (e.g.~\citealt{karakas14,delgadoinglada15}). An empirical study by~\citet{delgadoinglada15} has also shown that oxygen enrichment can occur in Galactic PNe with carbonaceous dust. It has been proposed by~\citet{garciahernandez16} that this can be explained by diffusive convective overshooting processes, in which core mixing is extended beyond the Schwarzschild boundary of main sequence stars~\citep{bohmvitense58,herwig97}, producing significant increases in oxygen abundances around sub-solar and solar metallicities~\citep{marigo01,pignatari16}.

While most abundances from observational studies have been measured using optical spectra, there are some advantages to analysing PNe using infrared spectra. These are described in several studies (e.g.~\citealt{rubin88,pottasch99,bernardsalas03}), but can be summarised as follows: extinction corrections are greatly reduced at IR wavelengths compared to those in the optical and ultraviolet regions; many ionic emission lines are observable for Ne, S and Ar within this wavelength range, and hence the need for ionisation correction factors (ICFs) in calculating elemental abundances is reduced; as these IR lines also originate from energy levels close to the ground state, both the uncertainties in the electron temperatures of any ion measured at IR wavelengths and temperature fluctuations within the PN can have little effect on the overall abundances. For this study in particular, the extinction corrections are reduced further as we have analysed the Galactic anti-centre, a region with much less extinction than the bulge (e.g.~\citealt{pottasch15}).

The presence of the Galactic metallicity gradient was made clear in a sample of H~{\tiny II} regions by~\citet{shaver83} for nitrogen, oxygen, sulphur, and argon. Since then, it has been further studied not only in H~{\tiny II} regions (e.g.~\citealt{martinhernandez02,esteban17,fernandezmartin17}) but also in PNe (e.g.~\citealt{maciel99,pottasch06,maciel15}), young B-type stars (e.g.~\citealt{fitzsimmons92,rolleston00}), Cepheid variables (e.g.~\citealt{andrievsky02b,andrievsky02a,andrievsky02c,luck03,lemasle13,genovali15}), open clusters (e.g.~\citealt{friel95}) and young stars (e.g.~\citealt{magrini17}). 

While the presence of the metallicity gradient in the Galactic disk is agreed upon over Galactocentric distances ($R_g$) in the range 4--10~kpc, its continuation towards the anti-centre is debated. Studies of H~{\tiny II} regions~\citep{esteban17,fernandezmartin17} and B-type stars~\citep{smartt00} have found that there is little variation in the gradient far from the Galactic centre, yet a study of Cepheid variables from \cite{andrievsky02c} showed the gradient flattening with $R_g$. Samples of PNe have also been previously analysed in the anti-centre with conflicting results. ~\cite{costa04} showed that the oxygen abundances of a group of PNe, 8--15~kpc away from the Galactic centre, did not directly follow the gradient but instead flattened beyond 10~kpc. This has also been observed in the nearby spiral galaxies M31, M33, M81 and NGC~300~\citep{magrini16}. However, the sample of \citet{henry10} suggested that the gradient steepened beyond this distance.

Chemical evolution models of the Milky Way have predicted that the radial abudance gradient will flatten over time due to several factors, such as the death of massive stars, which causes the metallicity to increase over time~(e.g.~\citealt{minchev13}), and radial migration (e.g.~\citealt{minchev12,minchev14,veraciro14,kubryk15b}), in which the angular momentum from stars is redistributed, leading to the movement of stars from the Galactic disk and hence contributing to a flattening radial metallicity gradient within the disk (e.g.~\citealt{sellwood02}).

Investigations into the time evolution of the radial metallicity gradient have given varying results, with several studies of PNe finding an overall steepening with time (e.g.~\citealt{maciel99,chiappini01,stanghellini10,kubryk15b}), suggesting that the Galactic disk formed slowly~\citep{chiappini97}. However,~\citet{maciel03} showed the gradient flattening over time. Studies of open clusters and field stars have also given varying conclusions on this matter~\citep{anders17}. 

In this paper, we have derived the abundances of neon, sulphur, and argon in a sample of 23 PNe located towards the Galactic anti-centre using IR data in order to study the metallicity gradient beyond 10~kpc, and compared them to other IR spectroscopic samples from the Milky Way that were analysed in the same way.

The layout of this paper is as follows: In Sect.~\ref{sec:Data} we discuss the source selection and the basic data reduction and extraction methods. In Sect.~\ref{sec:Analysis} we explain the methods used to calculate flux, intensity and abundance values as well as Galactocentric distances. The implications of these data on the metallicity gradient in the further regions of the Milky Way are considered in Sect.~\ref{sec:Discussion}. Finally, in Sect.~\ref{sec:Summary} we present our conclusions and summarise our main results.

\section{Data}  \label{sec:Data}

\subsection{Observations}
The observations were made with the Infrared Spectrograph on board the \emph{Spitzer Space Telescope} (Spitzer IRS)~\citep{werner04,houck04} through GTO programme 40035 (PI: J.~Bernard-Salas). The observations were carried out between December 2007 and December 2008 with the staring mode of the IRS using the short-low (SL), short-high (SH) and long-high (LH) modules, each allowing for simultaneous observations from two nod positions, at 1/3 and 2/3 of the way along the observing slit. These produced spectra with resolutions of R~$\sim$~60--127 in the range 5.2--14.5~$\mu$m, and of R~$\sim$~600 in the range 9.9--36.4~$\mu$m. 

The sources are listed in Table~\ref{tab:23PNe}. These were chosen according to the following two criteria: (a) the sources were located in the direction of the anti-centre ($l$~$=$~$120$--$240\degree$, $b$~$=$~$0\degree$~$\pm$~$20\degree$); (b) the physical sizes of the PNe were generally small enough to fit in the widest observing slit of \emph{Spitzer} IRS (LH, $11.1\arcsec$~$\times$~$22.3\arcsec$), thus minimising the aperture corrections required to account for the different slit sizes.

\subsection{Data reduction and extraction}
The basic calibration data (\emph{bcd}) image files obtained from the \emph{Spitzer} IRS were processed through the \emph{Spitzer} Science Centre (SSC) pipeline, version S18.18, then reduced and analysed through the Spectroscopic Modelling Analysis and Reduction Tool (\emph{SMART})~\citep{higdon04}. Rogue pixels were removed using the \emph{IRSCLEAN}\footnote{Available from the SSC website: http://ssc.spitzer.caltech.edu.} package. 

The methods of spectral extraction were based on the diameters of the sources (see Table~\ref{tab:scaling}). For the low-resolution SL module, sources with diameters $\leqslant~3\arcsec$ were extracted using the Advanced Optimal Extraction (\emph{AdOpt}) package~\citep{lebouteiller10}. This method weights each pixel based on their signal-to-noise ratios and is better suited to smaller sources. Those with diameters in the range $3\arcsec$--$8\arcsec$ were extracted using tapered column extractions, as the SL slit is not wide enough ($3.6\arcsec~\times~57\arcsec$) to detect all the flux in partially extended sources. The four PNe with diameters $\geqslant~10\arcsec$ were extracted using fixed column extractions as the FWHM of their emission was beyond the point spread function of the objects by factors up to approximately three. For the high-resolution SH and LH modules, full aperture extractions were used in each case; these weight all pixels equally from the aperture, allowing for most of the flux to be obtained from more extended sources if they covered an area larger than the observing slits. The flux values of a PN with a physical size bordering on two of these ranges do not change significantly ($\lesssim$~10\%).

\begin{figure*}[ht!]
\centering
\includegraphics[width=\textwidth]{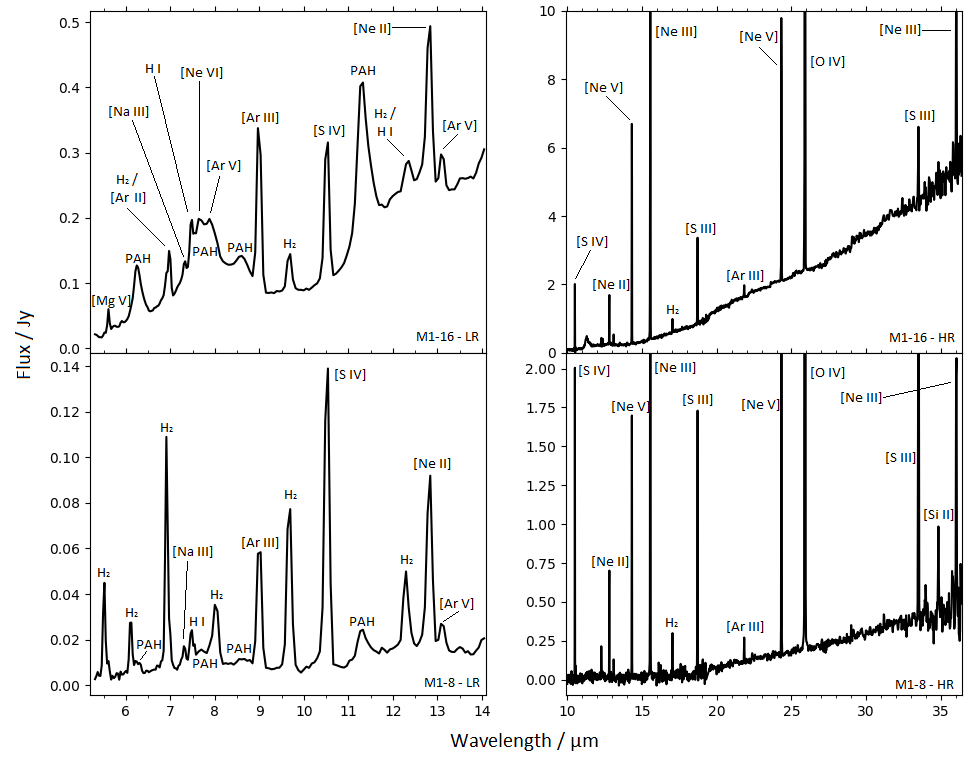}
\caption{Full \emph{Spitzer} IRS spectrum of M1-16 (top) and M1-8 (bottom). The low resolution spectra (SL) are shown on the left, and the high resolution spectra (SH and LH) are on the right.}
\label{fig:spec}
\end{figure*}

Most of the PNe in our sample were chosen such that all of the source flux would fall within the LH module. In fact, 15 of our 23 targets also have a physical diameter of $\leqslant~5\arcsec$, so in these cases most of the flux would also be detected by the smaller and narrower SH module ($4.7\arcsec~\times~11.3\arcsec$). Some of the source flux might still not be detected by the SL and SH modules, resulting in jumps in the baseline continuum. To account for this, we scale the SL and SH flux values by matching the continua in the overlapping wavelength regions. The scale factors are listed in Table~\ref{tab:scaling}. The full low- and high-resolution spectra of two representative PNe in the sample, M1-16 and M1-8, are shown in Figure~\ref{fig:spec}.

\begin{table}
\centering
\caption{Aperture corrections applied to each of the line flux values. A value of one implies that no correction is needed.}
\label{tab:scaling}
\begin{tabular}{ c c c c }
\hline\hline
PN & Diameter (\arcsec) & SL~$\rightarrow$~SH & SH~$\rightarrow$~LH \\
\hline
J320 & 7 & 1.40 & 1.47 \\
K3-65 & 5 & 1.20 & 1.15 \\
K3-66 & 2 & 1.05 & 1.00 \\
K3-67 & 2 & 1.00 & 1.00 \\
K3-68 & 12 & 1.50 & 2.00 \\
K3-69 & $<$1 & 1.00 & 1.00 \\
K3-70 & 2 & 1.00 & 1.05 \\
K3-71 & 3 & 1.00 & 1.00 \\
K3-90 & 10 & 1.00 & 2.00 \\
K4-48 & 2 & 1.15 & 1.08 \\
M1-1 & 5 & 1.00 & 1.50 \\
M1-6 & 4 & 1.20 & 1.12 \\
M1-7 & 11 & 1.00 & 2.20 \\
M1-8 & 18 & 1.00 & 2.00 \\
M1-9 & 3 & 1.12 & 1.07 \\
M1-14 & 5 & 1.50 & 1.30 \\
M1-16 & 3.6 & 1.38 & 1.18 \\
M1-17 & 3 & 1.18 & 1.05 \\
M2-2 & 7 & 2.30 & 2.00 \\
M3-2 & 8 & 1.00 & 1.00 \\
M4-18 & 4 & 1.12 & 1.00 \\
SaSt2-3 & $<$1 & 1.00 & 1.00 \\
Y-C 2-5 & 8 & 1.00 & 1.70 \\
\hline
\end{tabular}
\end{table}

\section{Analysis}  \label{sec:Analysis}

\subsection{Ionic abundances}
The ionic abundances obtained from the spectra have been calculated using the equation

\begin{equation}
\frac{n_{ion}}{n_{p}} = n_{e} \frac{I_{ion}}{I_{H\beta}} \frac{\lambda_{ul}}{\lambda_{H\beta}} \frac{\alpha_{H\beta}}{A_{ul}} \Bigg(\frac{n_{u}}{n_{ion}}\Bigg)^{-1}
\label{eq:abundance}
,\end{equation}

\noindent where $n_p$ is the proton density, $n_e$ is the electron density, $I_{ion}$ is the intensity of the ion, $\lambda_{ul}$ is the line wavelength, $\alpha_{H\beta}$ is the effective recombination coefficient for $H\beta$, $A_{ul}$ is the Einstein coefficient of spontaneous emission, and $n_{u}/n_{ion}$ is the ratio of the upper level population of the transition to the entire population of the ion. 

For the derivation of the neon, sulphur, and argon abundances, we have observed the emission lines for the most populated ions of these elements: Ne$^{+}$, Ne$^{2+}$, Ne$^{4+}$, S$^{2+}$, S$^{3+}$, Ar$^{+}$, Ar$^{2+}$ and Ar$^{4+}$. We complemented our IR ionic abundances with those of the missing ionisation states from optical spectra (Ne$^{3+}$, S$^{+}$ and Ar$^{3+}$) in the literature in order to avoid or reduce the need for ICFs. These values were primarily taken from~\citet{henry10} and references included in~\citet{sterling08}. We compared IR and optical data, so homogeneity in the slit sizes was assumed. This is a reasonable assumption as~\citet{henry10} use data from the Apache Point Observatory (APO), for which the slit size is 2$\arcsec$~$\times$~360$\arcsec$, and most of the PNe they have observed from this sample are $\lesssim$~4$\arcsec$ in diameter, so most of the flux will have been detected from these sources. We also considered S$^{4+}$, which is not directly detectable in IR or optical spectra, although it is only expected to contribute to the overall sulphur abundance for PNe with high IP values. Where there are no flux values available for any particular line in a PN, we applied correction factors (see~Sect.~\ref{sec:EAbuns}).

\subsection{Line flux measurements}
Flux values for the fine structure ionic emission lines with a $\geqslant$~3$\sigma$ detection were determined by applying Gaussian fits through the ISAP line fitting programme~\citep{sturm98} in \emph{SMART}. The raw $F(\lambda)$ values were calculated for each line in each of the two oberving positions (nods), which typically agreed to within 10\%. These were then averaged; in the few cases where there were any remaining low-level glitches obstructing a particular line, these were discarded in favour of the flux value from the nod without glitches. The associated uncertainties were propagated from those calculated from the individual flux measurements within each nod, unless the difference in flux between the two nods was greater than the assumed uncertainties. In this latter case, we considered the flux difference to be more representative of the uncertainty. The data were then corrected for extinction using the extinction law from~\citet{fluks94}. Table~\ref{tab:Flux} shows all values for the extinction corrected intensities, $I(\lambda)$. Upper limits of 3$\sigma$ or more were calculated for emission lines of the most important ions of neon, sulphur, and argon when there were no clear detections. 

\begin{sidewaystable*}
\centering
\caption{Selected line intensity values from the Galactic anti-centre PNe, in units of $10^{-14}$~erg~cm$^{-2}$~s$^{-1}$.}
\label{tab:Flux} 
\begin{tabular}{ c c c c c c c c c c c c c c c c c c }
\hline\hline
PN & \multicolumn{2}{c}{H {\tiny I}} & [Ne {\tiny II}] & \multicolumn{2}{c}{[Ne {\tiny III}]} & \multicolumn{2}{c}{[Ne {\tiny V}]} & \multicolumn{2}{c}{[S {\tiny III}]} & [S {\tiny IV}] & [Ar {\tiny II}] & \multicolumn{2}{c}{[Ar {\tiny III}]} & \multicolumn{2}{c}{[Ar {\tiny V}]} & [O {\tiny IV}] \\
 & \emph{7.5$\mu$m} & \emph{12.4$\mu$m} & \emph{12.8$\mu$m} & \emph{15.6$\mu$m} & \emph{36.0$\mu$m} & \emph{14.3$\mu$m} & \emph{24.3$\mu$m} & \emph{18.7$\mu$m} &\emph{33.5$\mu$m} & \emph{10.5$\mu$m} & \emph{7.0$\mu$m} & \emph{9.0$\mu$m} & \emph{21.8$\mu$m} & \emph{7.9$\mu$m} & \emph{13.1$\mu$m} & \emph{25.9$\mu$m} \\
\hline
J320 & 21.3 & 7.82 & 6.60 & 761 & 50.9 & ... & ... & 82.1 & 34.4 & 1020 & $<$1.72 & 28.0 & $<$2.39 & ... & ... & ... \\
K3-65 & 2.81$^{\ddagger}$ & ... & 15.0 & 186 & $<$24.5 & ... & ... & 49.8 & 38.4 & 53.7 & 8.31 & 17.9 & $<$1.81 & ... & ... & ... \\
K3-66 & 16.4$^{\dagger}$ & 6.40 & 94.7 & 185 & ... & $<$4.75 & $<$3.74 & 51.9 & 22.2 & 30.2 & 18.1$^{\dagger}$ & 32.7 & $<$2.82 & ... & ... & 3.20 \\
K3-67 & 32.7 & 11.7 & 28.3 & 778 & 54.2 & $<$5.12 & $<$3.73 & 108 & 42.1 & 519$^{\dagger}$ & 8.89 & 66.2 & 5.00$^{\dagger}$ & ... & ... & 7.16 \\
K3-68 & ... & ... & $<$2.95 & 37.4 & ... & 23.0$^{\dagger}$ & 23.6 & 10.6 & 11.1 & 98.8$^{\dagger}$ & $<$5.60 & $<$4.80 & $<$1.02 & $<$30.9$^{\ddagger}$ & 4.30 & 1190 \\
K3-69 & 4.88$^{\dagger}$ & ... & 5.72 & 176 & ... & 33.4 & 8.52 & 8.74$^{\dagger}$ & $<$10.4 & 20.7$^{\dagger}$ & $<$8.57 & 17.6$^{\dagger}$ & $<$3.04 & ... & ... & 35.3 \\
K3-70 & 3.56$^{\dagger}$ & ... & 5.44$^{\dagger}$ & 92.7 & ... & 23.4 & 15.4 & 28.8 & 16.9$^{\dagger}$ & 39.5 & 2.96 & 11.6$^{\dagger}$ & $<$1.48 & ... & ... & 61.0 \\
K3-71 & 1.33$^{\ddagger}$ & ... & $<$2.12 & 18.6 & ... & 32.6 & 43.4 & 3.87$^{\dagger}$ & 8.42 & 57.2 & $<$1.13 & 1.69$^{\ddagger}$ & $<$1.34 & 3.16$^{\ddagger}$ & 2.64 & 913 \\
K3-90 & ... & 1.97$^{\dagger}$ & $<$2.55 & 32.6 & ... & 131 & 152 & 6.03$^{\dagger}$ & 7.07$^{\ddagger}$ & 170 & $<$1.64 & 4.86$^{\dagger}$ & $<$1.41 & 9.13 & 15.7$^{\dagger}$ & 2070 \\
K4-48 & ... & 3.94$^{\dagger}$ & 20.0 & 497 & 23.0 & 22.4 & 8.02 & 38.2 & 9.22$^{\dagger}$ & 56.4 & 9.91 & 45.8$^{\dagger}$ & 2.52$^{\dagger}$ & ... & ... & 143 \\
M1-1 & ... & ... & $<$2.59 & 30.5 & ... & 670 & 753 & 14.3 & 10.6 & 210 & $<$2.03 & 4.61 & $<$1.32 & 6.86$^{\dagger}$ & 27.6 & 1620 \\
M1-6 & 104 & 36.2$^{\dagger}$ & 1320 & 57.5 & ... & ... & ... & 222 & 42.5$^{\dagger}$ & $<$9.38 & 298 & 151 & $<$10.5 & ... & ... & ... \\
M1-7 & 20.5 & 8.40$^{\dagger}$ & 61.4 & 874 & 52.7 & 2.89$^{\dagger}$ & 2.63 & 151 & 87.3 & 249 & 24.5$^{\dagger}$ & 77.7 & 6.10 & ... & ... & 371 \\
M1-8 & 6.31$^{\dagger}$ & ... & 28.9 & 323 & 26.4$^{\dagger}$ & 63.9 & 55.9 & 47.4 & 56.7 & 98.8 & * & 27.9 & 3.92 & $<$6.00 & 6.51 & 834 \\
M1-9 & 38.5$^{\dagger}$ & 8.94 & 111 & 371 & ... & ... & ... & 129 & 40.9 & 59.4 & 29.7$^{\dagger}$ & 49.7$^{\dagger}$ & $<$3.73 & ... & ... & ... \\
M1-14 & 73.1 & 25.9 & 727 & 446 & 37.7 & ... & ... & 375 & 112 & 41.3 & 39.2 & 170 & 10.9 & ... & ... & ... \\
M1-16 & 20.3 & 8.68 & 61.7 & 1070 & 76.4 & 256 & 147 & 68.7 & 32.1 & 91.3 & 43.2$^{\dagger}$ & 114$^{\dagger}$ & 8.09 & $<$14.6 & 12.1$^{\dagger}$ & 854 \\
M1-17 & 42.9 & 9.84$^{\dagger}$ & 47.2 & 849 & 67.3$^{\dagger}$ & 20.4 & 11.8 & 203 & 54.9 & 346 & 28.3 & 87.1$^{\dagger}$ & 5.89$^{\dagger}$ & ... & ... & 336 \\
M2-2 & 63.2 & 20.9$^{\dagger}$ & 7.86 & 1520 & 99.6 & $<$3.78 & $<$2.85 & 112 & 74.7 & 1250 & 2.22$^{\ddagger}$ & 77.1 & 6.27 & ... & ... & 216 \\
M3-2 & ... & ... & 2.59 & 10.4 & ... & 5.53 & 10.7$^{\dagger}$ & 3.59$^{\ddagger}$ & 6.33$^{\dagger}$ & 5.46$^{\dagger}$ & * & 1.93$^{\dagger}$ & $<$1.05 & ... & ... & 27.8 \\
M4-18 & ... & ... & 334 & $<$142 & ... & ... & ... & 14.8$^{\ddagger}$ & $<$13.5 & ... & 110 & ... & ... & ... & ... & ... \\
SaSt2-3 & 5.56$^{\dagger}$ & ... & 32.4 & $<$1.78 & ... & ... & ... & 7.12 & 7.11$^{\dagger}$ & ... & 32.4$^{\dagger}$ & ... & ... & ... & ... & ... \\
Y-C 2-5 & 2.62$^{\dagger}$ & ... & $<$1.36 & 86.9 & 25.9$^{\dagger}$ & ... & ... & 3.67 & $<$4.77 & 87.5 & * & 3.16 & $<$0.88 & ... & ... & 1190 \\
\hline
\end{tabular}
\tablefoot{Uncertainties are $<$~10\%, except: $^{\dagger}$ uncertainties 10--20\%, $^{\ddagger}$ uncertainties 20--35\%. * No line is observed, and the presence of the H$_2$ 0--0 S(5) feature at 6.9$\mu$m prevents an accurate upper limit being taken.}
\end{sidewaystable*}

\begin{table*}
\caption{$I(H\beta)$ values for the sample, in erg cm$^{-2}$ s$^{-1}$.}
\label{tab:Hbeta}
\centering
\begin{tabular}{ c c c c c }
\hline\hline
PN & $C(H\beta)$ & IR log $I(H\beta)$ & Lit. log $F(H\beta)$ & Lit. log $I(H\beta)$ \\
\hline
J320 & 0.24 & $-11.13$ & $-11.63^{(1)}$, $-11.39^{(2)}$ & $-11.15$ \\
K3-65 & 1.83$^{\dagger}$ & $-12.05$ & $-14.24$:$^{(1)}$ & $-12.41$: \\
K3-66 & 0.98 & $-11.26$ & $-12.22^{(3)}$ & $-11.24$ \\
K3-67 & 1.02 & $-10.91$ & $-12.13^{(1)}, -12.07^{(3)}$ & $-11.05$ \\
K3-68 & 0.80$^{\dagger}$ & ... & $-12.90^{(V)}$ & $-12.10$ \\
K3-69 & 1.34$^{\dagger}$ & $-11.80$ & $-13.25^{(1)}$ & $-11.91$ \\
K3-70 & 1.45 & $-11.92$ & $-13.54^{(1)}, -13.59^{(3)}$ & $-12.09$ \\
K3-71 & 1.14$^{\dagger}$ & $-12.35$ & $-13.62^{(1)}$ & $-12.48$ \\
K3-90 & 1.02 & $-11.70$ & $-13.40^{(3)}$ & $-12.38$ \\
K4-48 & 1.47 & $-11.39$ & $-12.93^{(3)}, -12.82^{(4)}$ & $-11.46$ \\
M1-1 & 0.6$^{\ddagger}$ & ... & $-11.84^{(1)}$, $-11.88^{(5)}$ & $-11.24$ \\
M1-6 & 1.57 & $-10.49$ & $-12.28^{(1)}, -12.34^{(3)}$ & $-10.77$ \\
M1-7 & 0.40 & $-11.15$ & $-12.21^{(1)}$, $-12.20^{(3)}$ & $-11.80$ \\
M1-8 & 1.1$^{\ddagger}$ & $-11.69$ & $-13.12^{(1)}$, $-12.37^{(5)}$ & $-12.02$ \\
M1-9 & 0.46 & $-10.99$ & $-11.66^{(3)}$, $-11.73^{(4)}$ & $-11.20$ \\
M1-14 & 0.69 & $-10.63$ & $-11.58^{(1)}, -12.20^{(3)}$ & $-10.89$ \\
M1-16 & 0.59 & $-11.12$ & $-12.80^{(3)}$, $-11.99^{(6)}$ & $-11.40$ \\
M1-17 & 0.96 & $-10.94$ & $-12.00^{(1)}$, $-11.89^{(7)}$ & $-10.93$ \\ 
M2-2 & 1.26 & $-10.69$ & $-12.22^{(1)}$, $-12.63^{(3)}$  & $-10.96$ \\
M3-2 & 0.22$^{\#}$ & ... & $-13.26^{(1)}$, $-12.32^{(5)}$ & $-12.10$ \\
M4-18 & 0.77 & ... & $-12.01^{(3)}$, $-12.15^{(8)}$ & $-11.24$ \\
SaSt2-3 & 0.73$^{\#}$ & $-11.77$ & $-12.68^{(9)}$ & $-11.95$ \\
Y-C 2-5 & 0.00 & $-12.07$ & $-12.65^{(3)}$, $-12.26^{(4)}$ & $-12.26$ \\
\hline 
\end{tabular}
\tablebib{(1) \citet{acker91}; (2) \citet{milingo02}; (3) \citet{henry10}; (4) \citet{cuisinier96}; (5) \citet{carrasco83}; (6) \citet{perinotto98}; (7) \citet{costa04}; (8) \citet{demarco99}; (9) \citet{pereira07}; (V) VizieR catalogue, given reference unverified.} 
\tablefoot{  : High errors. $C(H\beta)$ obtained from~\cite{henry10} except: $^{\dagger}$~from~\citet{giammanco11}; $^{\ddagger}$~from~\citet{condon98}; $^{\#}$~from~\citet{frew13}.}
\end{table*}

\begin{table}
\centering
\caption{Electron density values of PNe (cm$^{-3}$).}
\label{tab:ne}
\begin{tabular}{ c c c c }
\hline\hline
PN & $n_e$ (This work) & $n_e$ (Lit.) & Sources \\
\hline 
J320 & 3350~$\pm$~600 & 4800 & 1,2,3,4 \\
K3-65 & 1150~$\pm$~200 & ... & ... \\
K3-66 & 3150~$\pm$~500 & 7700 & 1,5 \\
K3-67 & 3900~$\pm$~550 & 4400 & 5,6,7,8 \\
K3-68 & 600~$\pm$~300 & 500 & 6 \\
K3-69 & \emph{3700~$\pm$~3000}$^{\ddagger}$ & ... & ... \\
K3-70 & 2000~$\pm$~650 & 2250 & 2,5,9 \\
K3-71 & 10000~$\pm$~2000$^{\dagger}$ & 10000 & 8 \\
K3-90 & 400~$\pm$~300 & 20000* & 5 \\
K4-48 & 8100~$\pm$~2250 & 2600 & 5,10 \\
M1-1 & 1300~$\pm$~450 & 4100 & 1,11 \\
M1-6 & 11450~$\pm$~4700 & 8500 & 2,5 \\
M1-7 & 1900~$\pm$~200 & 1050 & 2,5 \\
M1-8 & 350~$\pm$~150 & 440 & 2 \\
M1-9 & 5050~$\pm$~900 & 4600 & 2,5,10 \\
M1-14 & 5450~$\pm$~450 & 5400 & 2,5,12 \\
M1-16 & 2800~$\pm$~550 & 2300 & 2,5,10,13 \\
M1-17 & 6450~$\pm$~300 & 5000 & 2,9,14,15 \\
M2-2 & 1550~$\pm$~300 & 1600 & 5 \\
M3-2 & 230$^{\dagger}$ & 230 & 2 \\
M4-18 & 8000~$\pm$~3000$^{\dagger}$ & 8000 & 16,17 \\
SaSt2-3 & 600~$\pm$~450 & 2400 & 9 \\
Y-C 2-5 & \emph{3700~$\pm$~3000}$^{\ddagger}$ & ... & 2,5,10 \\
\hline
\end{tabular}
\tablebib{(1)~\citet{aller83}; (2)~\citet{costa04}; (3)~\citet{koeppen91}; (4)~\citet{milingo02}; (5)~\citet{henry10}; (6)~\citet{aller87}; (7)~\citet{kingsburgh94}; (8)~\citet{tamura87}; (9)~\citet{aksaker15}; (10)~\citet{cuisinier96}; (11)~\citet{aller86}; (12)~\citet{costa96}; (13)~\citet{perinotto98}; (14)~\citet{defreitas91}; (15)~\citet{peimbert95}; (16)~\citet{demarco99}; (17)~\citet{goodrich85}.}
\tablefoot{[S {\tiny III}] densities used when applicable. Literature values are averaged when there are multiple sources. No uncertainty was given for the literature value of M3-2. *~High density limit. $^{\dagger}$~Value from literature. $^{\ddagger}$~Mean $n_e$ value from those derived from [S~{\tiny III}] in our sample, with the standard deviation of all other data points as the associated uncertainty. }
\end{table}

\begin{table}
\centering
\caption{Electron temperature values of PNe as averages of literature values.}
\label{tab:Te}
\begin{tabular}{ c c c }
\hline\hline
PN & Ionic lines & $T_e$ (K) \\
\hline
J320 & [N~{\tiny II}], [O~{\tiny III}] & 11900~$\pm$~2300 \\
K3-65 & ... & \emph{11900~$\pm$~2600} \\
K3-66 & [N~{\tiny II}], [O~{\tiny II}], [O~{\tiny III}], [S~{\tiny II}], [S~{\tiny III}] & 10800~$\pm$~2500 \\
K3-67 & [N~{\tiny II}], [O~{\tiny II}], [O~{\tiny III}], [S~{\tiny III}] & 14400~$\pm$~3200 \\
K3-68 & [N~{\tiny II}], [O~{\tiny III}] & 19600~$\pm$~2000 \\
K3-69 & ...& \emph{11900~$\pm$~2600} \\
K3-70 & [N~{\tiny II}], [O~{\tiny II}], [O~{\tiny III}], [S~{\tiny III}] & 13700~$\pm$~4500 \\
K3-71 & [O~{\tiny III}] & 12600~$\pm$~2000 \\
K3-90 & [O~{\tiny II}], [O~{\tiny III}] & 12000~$\pm$~2500 \\
K4-48 & [N~{\tiny II}], [O~{\tiny II}], [O~{\tiny III}], [S~{\tiny II}], [S~{\tiny III}] & 12700~$\pm$~2300 \\
M1-1 & [O~{\tiny III}] & 14900~$\pm$~1500 \\
M1-6 & [N~{\tiny II}], [O~{\tiny III}], [S~{\tiny III}] & 9800~$\pm$~1900 \\
M1-7 & [N~{\tiny II}], [O~{\tiny II}], [O~{\tiny III}], [S~{\tiny II}], [S~{\tiny III}] & 10600~$\pm$~4200 \\
M1-8 & [N~{\tiny II}], [O~{\tiny III}] & 12900~$\pm$~1900 \\
M1-9 & [N~{\tiny II}], [O~{\tiny II}], [O~{\tiny III}], [S~{\tiny III}] & 10800~$\pm$~1800 \\
M1-14 & [N~{\tiny II}], [O~{\tiny II}], [O~{\tiny III}], [S~{\tiny III}] & 10000~$\pm$~3700 \\
M1-16 & [N~{\tiny II}], [O~{\tiny II}], [O~{\tiny III}], [S~{\tiny III}] & 11700~$\pm$~3000 \\
M1-17 & [N~{\tiny II}], [O~{\tiny III}] & 10700~$\pm$~2600 \\
M2-2 & [N~{\tiny II}], [O~{\tiny III}], [S~{\tiny III}] & 12500~$\pm$~1500 \\
M3-2 & [N~{\tiny II}] & 10200~$\pm$~1000 \\
M4-18 & [N~{\tiny II}], [O~{\tiny II}], [S~{\tiny II}] & 6100~$\pm$~3000 \\
SaSt2-3 & [N~{\tiny II}] & 9800~$\pm$~1400 \\
Y-C 2-5 & [N~{\tiny II}], [O~{\tiny II}], [O~{\tiny III}] & 13000~$\pm$~2400 \\
\hline
\end{tabular}
\tablefoot{K3-65 and K3-69 adopt the average temperature of the other sources due to lack of literature values; their associated uncertainties are given to be the standard deviation of all other values. Sources are the same as those given in Table~\ref{tab:ne}.}
\end{table}

\subsection{H$\beta$ intensity}
In the wavelength range of \emph{Spitzer} IRS we observed the emission features of several recombination transitions of atomic hydrogen (H~{\tiny I}), the strongest of which were observed at 7.5~$\mu$m and 12.4~$\mu$m. Both of these emission lines account for at least two transitions; the H~{\tiny I} 6--5, 8--6, 11--7 and 17--8 lines are blended around 7.5~$\mu$m (H~{\tiny I} 6--5 is the strongest of these transitions, contributing 74.43\% of the total flux) whereas the H~{\tiny I} 7--6 and 11--8 transitions both contribute to the emission line at 12.4~$\mu$m (H~{\tiny I} 7--6 provides 89.08\% of this flux). We applied the Balmer decrement to obtain values of $I(H\beta)$ from~\citet{hummer87}, interpolated to account for the electron density and temperature values of our PNe. When both of these IR emission features were observed, their $I(H\beta)$ values agreed by up to $\sim$~25\%, and an average was taken.  

Our calculated $I(H\beta)$ values are shown in Table~\ref{tab:Hbeta} alongside the $F(H\beta)$ and $I(H\beta)$ values from 4861\AA\ optical line measurements and the extinction coefficients, $C(H\beta)$, all taken from literature. In the four cases when neither of the two recombination lines were observed in a spectrum, we applied the Fluks extinction law \citep{fluks94} to these literature $F(H\beta)$ values. Our values agree with those in literature mostly to within a factor of two (more for K3-90 and M1-7). In these situations we favour our IR values, as the H~{\tiny I} lines are measured in the same spectra as the ionic emission that we have derived. Another advantage of using these lines to determine $F(H\beta)$ is that the extinction corrections are far smaller than those from optical wavelengths; $A_{\lambda}$~$<$~4.6 from the use of the 4861\AA\ $H\beta$ line, whereas from the IR recombination lines we find that $A_{\lambda}$~$<$~0.2.  

\subsection{Electron densities and temperatures}  \label{sec:neTe}
Both $n_e$ and $T_e$ are needed to determine abundance values; $n_e$ is a direct component of their calculation (see Equation~\ref{eq:abundance}) whereas $T_e$ designates the statistical populations of the excited electronic states present within the ion. These are listed in Tables~\ref{tab:ne} and \ref{tab:Te} respectively, and the transition probabilities and collision strengths used in calculating these values are shown in Table~\ref{tab:atomic}. These values were taken from TIPbase, part of the IRON project~\citep{hummer93}.

\begin{table*}
\caption{Atomic data for ions shown in Table~\ref{tab:Flux}.}
\label{tab:atomic}
\centering
\begin{tabular}{ c c c }
\hline\hline
Ion & Transition probability & Collision strength \\
\hline
Ne$^+$ & \citet{griffin01} & \citet{griffin01} \\
Ne$^{2+}$ & \citet{galavis97} & \citet{butler94} \\
Ne$^{4+}$ & \citet{galavis97} & \citet{griffin00} \\
Ne$^{5+}$ & \citet{mendoza83iau} & \citet{mitnik01} \\
S$^{2+}$ & \citet{mendoza82} & \citet{galavis95} \\
S$^{3+}$ & \citet{johnson86} & \citet{saraph99} \\
Ar$^+$ & \citet{mendoza83iau} & \citet{pelan95} \\
Ar$^{2+}$ & \citet{mendoza83} & \citet{galavis95} \\
Ar$^{4+}$ & \citet{mendoza82} & \citet{galavis95} \\
O$^{3+}$ & \citet{galavis98} & \citet{zhang94} \\
\hline
\end{tabular}
\end{table*}

Infrared lines originate from electronic transitions close to the ground state. Therefore, by analysing the ratio of $I(\lambda)$ values for two transitions of the same ion, we were able to obtain $n_e$ values that are mostly independent of temperature. Of the line flux ratios available from our spectra, we favoured those of the [S~{\tiny III}] 18.7~/~33.5~$\mu$m transitions as both of these lines are easy to measure in high resolution spectra and frequently seen together. While other line ratios were available in some PNe (e.g. [Ne~{\tiny III}], [Ne~{\tiny V}]), they were either detected in fewer of the PNe in our sample, or were detected at noisy wavelength regions. 
For instance, the [Ne~{\tiny III}] line at 36.0~$\mu$m is found at the upper wavelength region of the LH module, which is highly susceptible to noise above $\sim$~35~$\mu$m. The 14.3 and 24.3~$\mu$m lines of [Ne~{\tiny V}] were only observed in 12 of the 23 PNe in the sample, though the associated density values agree well with those of [S~{\tiny III}].  Uncertainties averaged $\sim$~20\% for values of $n_e$~$>$~1000~cm$^{-3}$, though this becomes larger for the few sources where $n_e$~$<$~1000~cm$^{-3}$. In the four cases where the two [S~{\tiny III}] lines were not directly measurable and any other line ratios were either not observed or affected by noise, we used values given in the literature from the [S~{\tiny II}] 6716~\AA\ /~6731~\AA\ line intensity ratio. We applied the mean value derived from the [S~{\tiny III}] line ratios in our sample of $n_e$~$=$~3700~cm$^{-3}$ for K3-69 and Y-C~2-5 as these lines were not observed in these PNe and there were no $n_e$ values given in literature. In these cases, while the uncertainty in density is high, the abundances are little affected, with neon and argon showing little change at the density extremes, and sulphur being affected by 20\% at most. All density values are shown in Table~\ref{tab:ne}, with the uncertainties reflecting those of the [S~{\tiny III}] 18.7~/~33.5~$\mu$m ratios. We note that for K3-90,~\citet{henry10} apply the high density limit to estimate $n_e$ despite having intensity values for the [S~{\tiny II}] 6716~\AA\ and 6731~\AA\ lines. This is due to the two values having high uncertainties. A density of $n_e$~$=$~800~cm$^{-3}$ would have been calculated with these values, which is almost within the error margins of our IR [S~{\tiny III}] line ratio.

Measurements of $T_e$ require electronic transitions with large differences in energy. For this study, we relied on temperatures calculated from optical line flux values based on the transition ratios of the [N~{\tiny II}], [O~{\tiny II}], [O~{\tiny III}], [S~{\tiny II}] and [S~{\tiny III}] lines calculated from the literature. As no $T_e$ values could be found in the literature for K3-65 and K3-69, for these two PNe we adopted the average value of $T_e$~$=$~11900~K. All of these values can be found in Table~\ref{tab:Te}.

\subsection{Elemental abundances}  \label{sec:EAbuns} 
One of the main advantages of analysing spectra at infrared wavelengths is that the main ionisation lines of neon, sulphur, and argon can be observed. From these lines, we have measured the ionic abundances of Ne$^{+}$, Ne$^{2+}$, Ne$^{4+}$, S$^{2+}$, S$^{3+}$, Ar$^{+}$, Ar$^{2+}$ and Ar$^{4+}$. We complemented these data with the ionic abundances of S$^{+}$ and Ar$^{3+}$ measured by~\citet{henry10} from optical spectra, hence fewer corrections are required in determining their elemental abundances. We accounted for Ne$^{3+}$ in sources with observable Ne$^{4+}$ emission, and we considered S$^{4+}$ in sources with O$^{3+}$, which has a greater IP (47.22~eV and 54.94~eV, respectively).

We corrected for these missing ionic abundances with ICFs. ICFs can either be determined empirically (e.g.~\citealt{surendiranath04,pottasch05}), by considering lines with similar IP values (e.g.~\citealt{peimbert69}), or from photoionisation models (e.g.~\citealt{natta80,kingsburgh94,kwitter01,delgadoinglada14a}). In many cases, argon and particularly sulphur are highlighted as being complicated to correct for, as the low IPs of higher ionisation states may lead to their greater contributions towards the overall elemental abundances. Many variants of the ICFs for these elements have been given in literature (e.g.~\citealt{kingsburgh94,thuan95,kwitter01}) with significant disagreement between some of them~\citep{vermeij02}.

To account for the abundances of missing ionisation states, we complemented our IR values with optical abundances derived by~\citet{henry10} where possible. To correct for S$^+$ and Ar$^{3+}$ when these values are not available, we calculated the percentage contributions of these ions towards their respective elemental abundances for the PNe in our sample with these ionic abundances and applied the mean values as ICFs. In each case, the minimum and maximum values were taken as the uncertainty limits. We also applied this method to account for Ar$^+$ in three sources for which the 7.0~$\mu$m line intensity cannot be measured. For Ne$^{3+}$ and S$^{4+}$, we considered the range of contributions of these ions to their respective elemental abundances as given by~\citet{bernardsalas08}, who calculated these from the analysis of the PN sample of~\citet{pottasch06} and also from the Galactic PN models of~\citet{surendiranath04} and~\citet{pottasch05}. Each of these ranges are shown in Table~\ref{tab:eicf}. We note that we only applied ICFs correcting for these missing ionic states in the PNe for which we observed other ions with greater or similar IP values.

\begin{table}
\centering
\caption{Percentage contributions of ions that have required the use of empirically calculated ICFs towards their respective elemental abundances.}
\label{tab:eicf}
\begin{tabular}{ c c c c }
\hline\hline
Ion & Range & Mean & Source \\
\hline
Ne$^{3+}$ & 2--33\% & 17.5\% & \citet{bernardsalas08} \\
S$^{+}$ & 1--20\% & 10\% & This work \\
S$^{4+}$ & 7--23\% & 15\% & \citet{bernardsalas08} \\
Ar$^{+}$ & 1--32\% & 13\% & This work \\
Ar$^{3+}$ & 3--46\% & 26\% & This work \\
\hline
\end{tabular}
\end{table}

Tables \ref{tab:NeAbuns}, \ref{tab:SAbuns}, and \ref{tab:ArAbuns} give the ionic and elemental abundance values for Ne, S and Ar respectively, with the empirical ICFs applied.  ICFs can be uncertain, so we also compared the resulting abundances calculated with our empirical ICFs with those calculated using well-established ICFs from the literature, such as those of~\citet{kingsburgh94}, \citet{kwitter01} and~\citet{delgadoinglada14a}. In all cases, we applied our ionic abundances. For the equations given in this section, we applied the notation A(X)~=~ICF(X$^{m+}$~$+$~X$^{n+}$)~$\times$~(X$^{m+}$~$+$~X$^{n+}$)/H, where A(X) is the elemental abundance of X. Table~\ref{tab:ICFsup} shows the supplementary data used in the following calculations from optical abundance studies. We note that some ICF equations from other studies use ionic ratios that are particularly sensitive to the electron density and temperature, such as those involving O$^+$ and O$^{2+}$. Hence, in cases where the [S~{\tiny III}] ion ratio does not adequately account for the electron density of an important ion for the ICF calculation, or the uncertainties in $T_e$ are large due to the dispersion of values over several studies, there may be additional uncertainty in the abundances calculated using these ICFs. 

\begin{table*}
\centering
\caption{Ionic and total abundances of neon ($\times 10^{-5}$). Optical values for ionic abundances were used for ions not observable in IR spectra.}
\label{tab:NeAbuns}
\begin{tabular}{ c c c c c c c c }
\hline\hline
Source & [Ne {\tiny II}] & [Ne {\tiny III}] & [Ne {\tiny V}] & [Ne {\tiny VI}] & ICF & Ne/H &  Ne/H lit. \\
 & {\tiny IP = 21.56eV} & {\tiny IP = 40.96eV} & {\tiny IP = 97.12eV} & {\tiny IP = 126.21eV} & & $\times 10^{-5}$ & $\times 10^{-5}$ \\
\hline 
J320 & 0.109 & 6.20 &  ... & ... & 1.00 & 6.3 $\pm$ 1.8 & \emph{5.3 $\pm$ 1.6} \\
K3-65 & 2.45 & 14.7 & ... & ... & 1.00 & 17.2 $\pm$ 5.9 & ... \\
K3-66 & 2.17 & 2.09 & $<$0.007 & ... & 1.00 & 4.3 $\pm$ 1.6 & 4.51 $\pm$ 1.79 \\
K3-67 & 0.253 & 3.47 & $<$0.003 & ... & 1.00 & 3.7 $\pm$ 1.1 & 3.79 $\pm$ 0.93, \emph{4.17 $\pm$ 1.46} \\
K3-68 & $<$0.338 & 2.11 & 0.218 & ... & 1.21 & 2.8 $\pm$ 1.6 & ... \\
K3-69 & 0.438 & 6.69 & 0.174 & ... & 1.21 & 8.9 $\pm$ 3.3 & ... \\
K3-70 & 0.507 & 4.23 & 0.151 & ... & 1.21 & 5.9 $\pm$ 2.9 & 7.01 $\pm$ 1.74 \\
K3-71 & $<$0.569 & 2.60 & 0.721 & ... & 1.21 & 4.0 $\pm$ 1.8 & ... \\
K3-90 & $<$0.155 & 0.952 & 0.500 & ... & 1.21 & 1.8 $\pm$ 0.8 & 14.1 $\pm$ 8.0 \\
K4-48 & 0.573 & 7.29 & 0.051 & ... & 1.21 & 9.6 $\pm$ 3.0 & 10.5 $\pm$ 2.5 \\
M1-1 & $<$0.048 & 0.275 & 0.886 & 0.014 & 1.21 & 1.4 $\pm$ 0.5 & \emph{8.9 $\pm$ 2.7}  \\
M1-6 & 5.65 & 0.129 & ... & ... & 1.00 & 5.8 $\pm$ 2.3 & 1.15 $\pm$ 1.53 \\
M1-7 & 1.06 & 7.32 & 0.003 & ... & 1.21 & 10.2 $\pm$ 6.0 & 20.8 $\pm$ 4.8 \\
M1-8 & 1.62 & 8.73 & 0.235 & ... & 1.21 & 12.8 $\pm$ 5.1 & ... \\
M1-9 & 1.39 & 2.32 & ... & ... & 1.00 & 3.7 $\pm$ 1.8 & 4.04 $\pm$ 1.18 \\
M1-14 & 4.18 & 1.28 & ... & ... & 1.00 & 5.5 $\pm$ 2.2 & 1.61 $\pm$ 0.45 \\
M1-16 & 0.991 & 8.44 & 0.270 & 0.008 & 1.21 & 11.8 $\pm$ 5.8 & 10.8 $\pm$ 2.6, \emph{7.0 $\pm$ 2.0} \\
M1-17 & 0.539 & 4.91 & 0.017 & ... & 1.21 & 6.6 $\pm$ 1.7 & ... \\
M2-2 & 0.045 & 4.24 & $<$0.001 & ... & 1.00 & 4.3 $\pm$ 1.0 & 4.89 $\pm$ 1.19, \emph{5.89 $\pm$ 3.53} \\
M3-2 & 0.426 & 0.819 & 0.052 & ... & 1.21 & 1.6 $\pm$ 0.9 & ... \\
M4-18 & 10.6 & $<$2.32 & ... & ... & 1.00 & 10.6 $\pm$ 3.9 & \emph{0.54 $\pm$ 0.06} \\
SaSt2-3 & 2.56 & $<$0.068 & ... & ... & 1.00 & 2.6 $\pm$ 1.5 & ... \\
Y-C 2-5 & $<$0.188 & 5.90 & $<$0.036 & ... & 1.00 & 5.9 $\pm$ 3.4 & ... \\
\hline
\end{tabular}
\tablefoot{ICFs applied for [Ne~{\tiny IV}] contributions - see Sect.~\ref{sec:EAbuns}.  Literature values from~\citet{henry10}. Literature abundances in italics are from sources given in~\citet{sterling08}.}
\end{table*}

\begin{table*}
\centering
\caption{Ionic and total abundances of sulphur ($\times 10^{-6}$). Optical values for ionic abundances were used for ions not observable in IR spectra.}
\label{tab:SAbuns}
\begin{tabular}{ c c c c c c c c c }
\hline\hline
Source & [S {\tiny II}] lit. & [S {\tiny III}] & [S {\tiny IV}] & ICF & S/H & S/H lit. \\
 & {\tiny IP = 10.36eV} & {\tiny IP = 23.34eV} & {\tiny IP = 34.79eV} & & $\times 10^{-6}$ & $\times 10^{-6}$ \\
\hline 
J320 & ... & 1.25 & 3.16 & 1.33 & 5.8 $\pm$ 1.7 & \emph{14 $\pm$ 4} \\
K3-65 & ... & 6.64 & 1.45 & 1.33 & 10.8 $\pm$ 3.7 & ... \\
K3-66 & 0.15 $\pm$ 0.10 & 1.13 & 0.126 & 1.18 & 1.7 $\pm$ 0.6 & 1.70 $\pm$ 0.55 \\
K3-67 & 0.10 $\pm$ 0.04 & 0.880 & 0.886 & 1.18 & 2.2 $\pm$ 0.7 & 2.04 $\pm$ 0.76, \emph{5.01 $\pm$ 1.75} \\
K3-68 & ... & 0.899 & 1.92 & 1.33 & 3.8 $\pm$ 2.1 & ... \\
K3-69 & ... & 0.629 & 0.302 & 1.33 & 1.2 $\pm$ 0.5 & ... \\
K3-70 & 0.30 $\pm$ 0.04$^{\ddagger}$ & 2.21 & 0.642 & 1.18 & 3.7 $\pm$ 1.8 & 3.44 $\pm$ 1.07 \\
K3-71 & ... & 1.37 & 3.73 & 1.33 & 6.8 $\pm$ 3.0 & ... \\
K3-90 & 0.36 $\pm$ 0.15 & 0.303 & 1.65 & 1.18 & 2.7 $\pm$ 1.2 & 1.86 $\pm$ 1.42 \\
K4-48 & 0.19 $\pm$ 0.06 & 1.32 & 0.370 & 1.18 & 2.2 $\pm$ 0.7 & 1.65 $\pm$ 0.51, \emph{1.95 $\pm$ 0.59} \\
M1-1 & ... & 0.207 & 0.656 & 1.33 & 1.2 $\pm$ 0.4 & ... \\
M1-6 & (0.48 $\pm$ 0.78) & 1.45 & $<$0.011 & 1.11 & 1.6 $\pm$ 0.6 & 2.66 $\pm$ 1.29, \emph{1.91 $\pm$ 0.57} \\
M1-7 & 0.75 $\pm$ 0.17 & 2.24 & 0.743 & 1.18 & 4.4 $\pm$ 2.6 & 3.88 $\pm$ 1.10 \\
M1-8 & ... & 2.17 & 0.888 & 1.33 & 4.1 $\pm$ 1.6 & ... \\
M1-9 & 0.16 $\pm$ 0.10 & 1.69 & 0.151 & 1.00 & 2.0 $\pm$ 1.0 & 2.10 $\pm$ 0.64, \emph{1.29 $\pm$ 0.39} \\
M1-14 & 0.14 $\pm$ 0.07 & 2.35 & 0.049 & 1.00 & 2.5 $\pm$ 1.0 & 2.19 $\pm$ 0.67, \emph{0.81 $\pm$ 0.24} \\
M1-16 & 0.24 $\pm$ 0.07 & 0.981 & 0.265 & 1.18 & 1.8 $\pm$ 0.9 & 1.55 $\pm$ 0.46, \emph{1.80 $\pm$ 0.50} \\
M1-17 & 0.37 $\pm$ 0.03$^{\ddagger}$ & 2.69 & 0.857 & 1.18 & 4.6 $\pm$ 1.2 & \emph{9.55 $\pm$ 2.87} \\
M2-2 & 0.01 $\pm$ 0.00 & 0.526 & 1.21 & 1.18 & 2.1 $\pm$ 0.5 & 1.10 $\pm$ 0.51 \\
M3-2 & ... & 0.512 & 0.141 & 1.33 & 0.87 $\pm$ 0.49 & ... \\
M4-18 & 2.89 $\pm$ 0.74$^{\dagger}$ & 0.665 & ... & 1.00 & 3.6 $\pm$ 1.6 & 3.56 $\pm$ 3.85, \emph{1.45 $\pm$ 0.20} \\
SaSt2-3 & 0.24 $\pm$ 0.04$^{\ddagger}$ & 0.476 & ... & 1.00 & 0.70 $\pm$ 0.42 & ... \\
Y-C 2-5 & ... & 0.462 & 2.27 & 1.33 & 3.6 $\pm$ 2.1 & ... \\
\hline
\end{tabular}
\tablefoot{ICFs applied to account for [S~{\tiny II}] and [S~{\tiny V}] contributions - see Sect.~\ref{sec:EAbuns}. Literature values from~\citet{henry10}. Literature abundances in italics are from sources given in~\citet{sterling08}. $^{\dagger}$~Abundance calculated from flux data in~\citet{demarco99}. $^{\ddagger}$~Abundances calculated from flux data in~\citet{aksaker15}.}
\end{table*}

\begin{table*}
\centering
\caption{Ionic and total abundances of argon ($\times 10^{-7}$). Optical values for ionic abundances were used for ions not observable in IR spectra.}
\label{tab:ArAbuns}
\begin{tabular}{ c c c c c c c c }
\hline\hline
Source & [Ar {\tiny II}] & [Ar {\tiny III}] & [Ar {\tiny IV}] lit. & [Ar {\tiny V}] & ICF & Ar/H & Ar/H lit. \\
 & {\tiny IP = 15.76eV} & {\tiny IP = 27.63eV} & {\tiny IP = 40.74eV} & {\tiny IP = 59.81eV} & & $\times 10^{-7}$ & $\times 10^{-7}$ \\
\hline 
J320 & $<$0.180 & 3.55 & ... & ... & 1.35 & 4.8 $\pm$ 1.4 & \emph{9.4 $\pm$ 2.8} \\
K3-65 & 8.45 & 22.0 & ... & ... & 1.35 & 41.2 $\pm$ 14.2 & ... \\
K3-66 & 2.66 & 5.64 & ... & ... & 1.35 & 11.2 $\pm$ 4.4 & 5.4 $\pm$ 1.1 \\
K3-67 & 0.487 & 4.48 & 1.73 $\pm$ 0.35 & ... & 1.00 & 6.7 $\pm$ 2.0 & 6.0 $\pm$ 1.2, \emph{10.0 $\pm$ 3.5} \\
K3-68 & $<$3.83 & $<$3.96 & ... & 1.05 & 1.35 & $<$6.7 $\pm$ 4.2 & ... \\
K3-69 & $<$4.13 & 10.3 & ... & ... & 1.35 & 13.9 $\pm$ 6.7 & ... \\
K3-70 & 1.70 & 8.17 & 3.16 $\pm$ 1.09 & ... & 1.00 & 13.0 $\pm$ 6.6 & 14.6 $\pm$ 3.0 \\
K3-71 & $<$1.89 & 3.59 & ... & 1.98 & 1.35 & 7.5 $\pm$ 3.6 & ... \\
K3-90 & $<$0.350 & 2.23 & 3.44 $\pm$ 0.64 & 1.78 & 1.00 & 7.5 $\pm$ 3.4 & 13.3 $\pm$ 2.4 \\
K4-48 & 1.78 & 10.2 & 4.21 $\pm$ 0.80 & ... & 1.00 & 16.2 $\pm$ 5.3 & 19.9 $\pm$ 3.7, \emph{5.4 $\pm$ 4.1} \\
M1-1 & $<$0.227 & 0.632 & ... & 1.05 & 1.64 & 2.8 $\pm$ 1.3 & \emph{21.0 $\pm$ 8.6} \\
M1-6 & 8.32 & 5.21 & ... & ... & 1.35 & 18.3 $\pm$ 7.2 & 9.1 $\pm$ 2.3, \emph{40.7 $\pm$ 12.2} \\
M1-7 & 2.69 & 10.2 & 3.27 $\pm$ 0.59 & ... & 1.00 & 16.2 $\pm$ 9.5 & 36.4 $\pm$ 6.4 \\
M1-8 & * & 11.7 & ... & 0.703 & 1.64 & 20.3 $\pm$ 10.6 & ... \\
M1-9 & 2.38 & 4.85 & 0.22 $\pm$ 0.08 & ... & 1.00 & 7.5 $\pm$ 3.7 & 8.0 $\pm$ 2.0, \emph{10.0 $\pm$ 3.0} \\
M1-14 & 1.46 & 7.67 & ... & ... & 1.35 & 12.3 $\pm$ 4.9 & 9.5 $\pm$ 2.0, \emph{20.4 $\pm$ 6.1} \\
M1-16 & 4.35 & 13.9 & 3.56 $\pm$ 0.68 & 0.400 & 1.00 & 22.2 $\pm$ 11.1 & 22.2 $\pm$ 4.0, \emph{18.0 $\pm$ 3.0} \\
M1-17 & 2.07 & 7.83 & ... & ... & 1.35 & 13.4 $\pm$ 3.9 & \emph{33.1 $\pm$ 9.9} \\
M2-2 & 0.079 & 3.32 & 3.00 $\pm$ 0.57 & ... & 1.35 & 8.7 $\pm$ 2.0 & 6.8 $\pm$ 1.2, \emph{8.9 $\pm$ 5.4} \\
M3-2 & * & 2.41 & ... & ... & 1.64 & 4.0 $\pm$ 2.5 & ... \\
M4-18 & 24.8 & ... & ... & ... & 1.00 & 24.8 $\pm$ 9.4 & ... \\
SaSt2-3 & 16.6 & ... & ... & ... & 1.00 & 16.6 $\pm$ 10.0 & ... \\
Y-C 2-5 & * & 3.30 & ... & ... & 1.64 & 5.4 $\pm$ 3.4 & ... \\
\hline
\end{tabular}
\tablefoot{ICFs applied to account for [Ar~{\tiny II}] and [Ar~{\tiny IV}] contributions - see Sect.~\ref{sec:EAbuns}. Literature values from~\citet{henry10}. Literature abundances in italics are from sources given in~\citet{sterling08}. * Upper limits for 6.99$\mu$m [Ar {\tiny II}] are inaccurate due to the contribution of the H$_2$ 0--0 S(5) line at 6.9$\mu$m.}
\end{table*}

\begin{table*}
\centering
\caption{Ionic and elemental abundances from optical data in the literature, used in the calculations of neon, sulphur, and argon ICFs in Tables~\ref{tab:NeICFs}--\ref{tab:ArICFs}.}
\label{tab:ICFsup}
\begin{tabular}{ c c c c c c c c }
\hline\hline
 & He$^+$/H$^+$ & He$^{2+}$/H$^+$ & N$^+$/H$^+$ & N/H & O$^+$/H$^+$ & O$^{2+}$/H$^+$ & O/H \\
PN & $\times$10$^{-3}$ & $\times$10$^{-3}$ & $\times$10$^{-6}$ & $\times$10$^{-6}$ & $\times$10$^{-5}$ & $\times$10$^{-5}$ & $\times$10$^{-5}$ \\
\hline 
J320 & ... & ... & \emph{0.23~$\pm$~0.08} & \emph{14.8~$\pm$~4.4} & \emph{0.42~$\pm$~0.13} & 25.7~$\pm$~7.7 & 27.5~$\pm$~8.3 \\
K3-66 & 88~$\pm$~11 & 0.32~$\pm$~0.10 & 12.6~$\pm$~4.1 & 34.1~$\pm$~8.1 & 59~$\pm$~38 & 10.0~$\pm$~2.2 & 16.0~$\pm$~4.8 \\
K3-67 & 93~$\pm$~14 & 0.24~$\pm$~0.04 & 3.6~$\pm$~0.9 & 79~$\pm$~23 & 5.9~$\pm$~1.8 & 12.5~$\pm$~3.2 & 13.1~$\pm$~3.3 \\
K3-70 & 98~$\pm$~15 & 21.5~$\pm$~3.2 & 52~$\pm$~12 & 305~$\pm$~81 & 20.4~$\pm$~5.6 & 7.8~$\pm$~2.0 & 12.0~$\pm$~2.7 \\
K3-90 & 4.9~$\pm$~2.5 & 105~$\pm$~17 & ... & ... & 1.3~$\pm$~0.7 & 2.7~$\pm$~0.7 & 62~$\pm$~33 \\
K4-48 & 107~$\pm$~14 & 17.3~$\pm$~2.6 & 22.5~$\pm$~5.4 & 206~$\pm$~50 & 35~$\pm$~11 & 24.3~$\pm$~5.8 & 32.3~$\pm$~7.1 \\
M1-6 & ... & ... & 42~$\pm$~20 & 58~$\pm$~26 & 241~$\pm$~87 & 9.5~$\pm$~2.9 & 34~$\pm$~11 \\
M1-7 & 110~$\pm$~14 & 15.7~$\pm$~2.4 & 61~$\pm$~14 & 252~$\pm$~70 & 115~$\pm$~30 & 29.9~$\pm$~6.8 & 47.3~$\pm$~9.2 \\
M1-9 & 104~$\pm$~14 & 0.11~$\pm$~0.04 & 8.8~$\pm$~2.8 & 37~$\pm$~13 & 47~$\pm$~31 & 14.5~$\pm$~3.3 & 19.2~$\pm$~5.0 \\
M1-14 & 96~$\pm$~12 & 0.10~$\pm$~0.05 & 14.0~$\pm$~3.9 & 46~$\pm$~13 & 89~$\pm$~47 & 20.1~$\pm$~4.4 & 29.0~$\pm$~6.9 \\
M1-16 & 105~$\pm$~15 & 25.2~$\pm$~3.8 & 81~$\pm$~20 & 543~$\pm$~145 & 50~$\pm$~14 & 21.8~$\pm$~5.3 & 33.2~$\pm$~7.2 \\
M2-2 & 104~$\pm$~14 & 7.8~$\pm$~1.2 & 0.36~$\pm$~0.12 & 15.8~$\pm$~6.6 & 4.5~$\pm$~0.8 & 17.9~$\pm$~4.3 & 19.8~$\pm$~4.7 \\
Y-C 2-5 & 39.7~$\pm$~5.9 & 62.4~$\pm$~9.2 & 0.15~$\pm$~0.05 & 55~$\pm$~34 & 0.79~$\pm$~0.39 & 11.1~$\pm$~2.9 & 28.6~$\pm$~7.7 \\
\hline
\end{tabular}
\tablefoot{All values have been taken from~\citet{henry10}, except for those of J320 which are from~\citet{costa04} with an assumed 30\% uncertainty applied. Values in italics are not involved in any future calculations, and have only been included for completeness.  }
\end{table*}

\subsubsection{Neon}   \label{sec:ne431}
At mid-IR wavelengths, lines of Ne$^{+}$, Ne$^{2+}$, Ne$^{4+}$ and Ne$^{5+}$ can be measured, though Ne$^{3+}$ is best observed in the optical and near-UV regions, respectively at $\sim$~4720$\AA$ and $\sim$~2424$\AA$. Unfortunately, no literature values exist for the abundance of Ne$^{3+}$ in any of the PNe in the anti-centre sample.

The photoionisation model of~\citet{kingsburgh94}, also used by~\citet{kwitter01}, used the following ICFs:

\begin{equation}
\textrm{ICF(Ne}^{2+} + \textrm{Ne}^{4+})~=~1.5
\label{eq:kbne24}
,\end{equation}
\begin{equation}
\textrm{ICF(Ne}^{2+})~=~\frac{\textrm{O}}{\textrm{O}^{2+}}
\label{eq:kbne2}
.\end{equation}

\noindent However, the abundances of sources with weak radiation fields are underestimated due to the disregard of the Ne$^{+}$ ionic contribution to the total neon abundance (e.g.~\citealt{tsamis13}). This problem is also observed in the ionic abundances of several sources in our sample of anti-centre PNe, in which Ne$^{+}$ sometimes contributes more to the overall elemental abundance than Ne$^{2+}$. These include K3-66, M1-6 and M1-14 (see Table~\ref{tab:NeAbuns}).

Recently,~\citet{delgadoinglada14a} produced a newer set of ICF models to account for parameters such as effective temperatures and stellar luminosities. For Ne/H corrections, they apply:

\begin{equation}
\textrm{ICF(Ne}^{2+} + \textrm{Ne}^{4+})~=~(1.31 + 12.68\nu^{2.57})^{0.27}
\label{eq:dine24}
\end{equation}

\noindent where $\nu$~=~He$^{2+}$/(He$^+$~+~He$^{2+}$). However, the requirement for Ne$^{4+}$ limits the usability of this correction. They also state that the ICFs will overestimate the neon abundances unless 0.4~$\lesssim$~$\nu$~$\lesssim$~0.6. The He$^+$ and He$^{2+}$ abundances given for 12 of the 23 PNe in the anti-centre sample from \citet{henry10} give $\nu$ values outside this range, ten of these sources having values of $\nu$~$<$~0.2 and 5 of these with $\nu$~$<$~0.005. PNe with very low $\nu$ values have small He$^{2+}$ ionic abundances, which typically indicates low-ionisation sources with little or no Ne$^{4+}$ emission. However, this is not true for Y-C~2-5, which has a relatively large $\nu$ (0.61) but no observable Ne$^{4+}$ emission in its \emph{Spitzer} IRS spectrum.

Table~\ref{tab:NeICFs} shows a comparison between the neon abundances calculated with both our empirical ICFs and the well-established ICFs of~\citet{kingsburgh94} and \citet{delgadoinglada14a}. There is good agreement in almost all cases between the two sets of values, though the disregard of Ne$^+$ leads to major underestimates from the \citet{kingsburgh94} model for M1-6 and M1-14. For Y-C~2-5, the applied ICF of 2.58 leads to a much greater abundance than predicted empirically. However, as the mid-IR spectrum of this PN shows the strong emission line of Ne$^{2+}$ but not those of Ne$^+$ or Ne$^{4+}$, it is possible that the Ne$^{3+}$ for which we are correcting may give a significant contribution to the neon abundance. The fact that we see large amounts of He$^{2+}$ compared to He$^+$ in this source shows that the radiation field in Y-C~2-5 is greater than 54.4~eV (the IP of He$^{2+}$), and as Ne$^{3+}$ ionises at 63.5~eV, a large ICF may be required.

\begin{table*}
\centering
\caption{Comparison of the neon abundances and ICFs used in this study with those in which ICFs from other sources have been applied (see Sect.~\ref{sec:ne431}).}
\label{tab:NeICFs}
\begin{tabular}{ c c c c c c c }
\hline\hline
 & \multicolumn{2}{c}{This work} & \multicolumn{2}{c}{KB94} & \multicolumn{2}{c}{DI14} \\
 & Ne/H & & Ne/H & & Ne/H & \\
PN & $\times 10^{-5}$ & ICF(Ne) & $\times 10^{-5}$ & ICF(Ne) & $\times 10^{-5}$ & ICF(Ne) \\
\hline 
J320 & 6.3~$\pm$~1.8$^{* \dagger}$ & 1.00 & 6.6~$\pm$~3.3$^{\dagger}$ & 1.07 & ... & ... \\
K3-66 & 4.3~$\pm$~1.6$^{* \dagger}$ & 1.00 & 3.4~$\pm$~1.8$^{\dagger}$ & 1.61 & ... & ... \\
K3-67 & 3.7~$\pm$~1.1$^{* \dagger}$ & 1.00 & 3.6~$\pm$~1.7$^{\dagger}$ & 1.05 & ... & ... \\
K3-68 & 2.8~$\pm$~1.6$^{* \dagger \ddagger}$ & 1.21 & 3.5~$\pm$~0.7$^{\dagger \ddagger}$ & 1.50 & ... & ... \\
K3-69 & 8.9~$\pm$~3.3$^{* \dagger \ddagger}$ & 1.21 & 10.3~$\pm$~3.3$^{\dagger \ddagger}$ & 1.50 & ... & ... \\
K3-70 & 5.9~$\pm$~2.9$^{* \dagger \ddagger}$ & 1.21 & 6.6~$\pm$~3.2$^{\dagger \ddagger}$ & 1.50 & 4.9~$\pm$~2.5$^{\dagger \ddagger}$ & 1.11 \\
K3-71 & 4.0~$\pm$~1.8$^{* \dagger \ddagger}$ & 1.21 & 5.0~$\pm$~2.2$^{\dagger \ddagger}$ & 1.50 & ... & ... \\
K3-90 & 1.8~$\pm$~0.8$^{* \dagger \ddagger}$ & 1.21 & 2.2~$\pm$~0.9$^{\dagger \ddagger}$ & 1.50 & 2.9~$\pm$~1.2$^{\dagger \ddagger}$ & 1.98  \\
K4-48 & 9.6~$\pm$~3.0$^{* \dagger \ddagger}$ & 1.21 & 11.0~$\pm$~3.1$^{\dagger \ddagger}$ & 1.50 & 8.0~$\pm$~3.0$^{\dagger \ddagger}$ & 1.09 \\
M1-1 & 1.4~$\pm$~0.5$^{* \dagger \ddagger \#}$ & 1.21 & 1.7~$\pm$~0.6$^{\dagger \ddagger}$ & 1.50 & ... & ... \\
M1-6 & 5.8~$\pm$~2.3$^{* \dagger}$ & 1.00 & 0.46~$\pm$~0.26$^{\dagger}$ & 3.54 & ... & ... \\
M1-7 & 10.2~$\pm$~6.0$^{* \dagger \ddagger}$ & 1.21 & 11.0~$\pm$~5.0$^{\dagger \ddagger}$ & 1.50 & 8.0~$\pm$~2.4$^{\dagger \ddagger}$ & 1.09 \\
M1-8 & 12.8~$\pm$~5.1$^{* \dagger \ddagger}$ & 1.21 & 13.4~$\pm$~3.2$^{\dagger \ddagger}$ & 1.50 & ... & ... \\
M1-9 & 3.7~$\pm$~1.8$^{* \dagger}$ & 1.00 & 3.1~$\pm$~1.3$^{\dagger}$ & 1.32 & ... & ... \\
M1-14 & 5.5~$\pm$~2.2$^{* \dagger}$ & 1.00 & 1.8~$\pm$~0.9$^{\dagger}$ & 1.44 & ... & ... \\
M1-16 & 11.8~$\pm$~5.8$^{* \dagger \ddagger \#}$ & 1.21 & 13.1~$\pm$~6.9$^{\dagger \ddagger}$ & 1.50 & 9.7~$\pm$~3.5$^{\dagger \ddagger}$ & 1.11 \\
M1-17 & 6.6~$\pm$~1.7$^{* \dagger \ddagger}$ & 1.21 & 7.4~$\pm$~2.3$^{\dagger \ddagger}$ & 1.50 & ... & ... \\
M2-2 & 4.3~$\pm$~1.0$^{* \dagger}$ & 1.00 & 4.7~$\pm$~1.9$^{\dagger}$ & 1.11 & 4.6~$\pm$~2.0$^{\dagger \ddagger}$ & 1.08 \\
M3-2 & 1.6~$\pm$~0.9$^{* \dagger \ddagger}$ & 1.21 & 1.3~$\pm$~0.5$^{\dagger \ddagger}$ & 1.50 & ... & ... \\
Y-C 2-5 & 5.9~$\pm$~3.4$^{* \dagger}$ & 1.00 & 15.2~$\pm$~6.9$^{\dagger}$ & 2.58 & ... & ... \\
\hline
\end{tabular}
\tablefoot{KB94~$=$~\citet{kingsburgh94}; DI14~$=$~\citet{delgadoinglada14a}. In all cases, ICFs are applied to our neon ionic abundances from Table~\ref{tab:NeAbuns}. Superscript symbols show the ions considered in the calculations: $^{*} =$~Ne$^+$; $^{\dagger} =$~Ne$^{2+}$; $^{\ddagger} =$~Ne$^{4+}$; $^{\#} =$~Ne$^{5+}$. M4-18 and SaSt2-3 are not included on this table as their Ne$^{2+}$ abundances are upper limits. K3-65 is not included as its helium and oxygen abundances have not been found in the literature.}
\end{table*}

\subsubsection{Sulphur}    \label{sec:s432}
As ionisation models are typically applied to optical spectra, it is normal to only see corrections for S$^{3+}$ from S$^+$ and S$^{2+}$ ionic abundance measurements. In these studies, the low IP of S$^{2+}$ (34.8~eV) has always been taken as an indication that larger ionic states are likely to be present in PNe, and the similar IP of O$^+$ (35.1~eV) is generally considered in obtaining an ICF for sulphur. \citet{dinerstein80} carried out an IR spectroscopic survey of 12 PNe and found that the commonly used ICF of O/O$^+$ can overpredict the measured abundances of S$^{3+}$. However, the presence of even greater ionisation states must also be considered.

Based on models of H~{\tiny II} regions from~\citet{stasinska78}, the ICF for sulphur from optical spectra was calculated by~\citet{barker80} to be

\begin{equation}
\textrm{ICF(S}^{+} + \textrm{S}^{2+}) = \bigg[1 - \bigg(1 - \frac{\textrm{O}^{+}}{\textrm{O}}\bigg)^{\alpha}\bigg]^{-1/\alpha}
\label{eq:kbs12}
\end{equation}

\noindent where $\alpha$~$=$~3, though subsequent studies argued that $\alpha$~$=$~2 \citep{french81} or 2~$\leqslant$~$\alpha$~$\leqslant$~3 \citep{garnett89} better represented the sulphur abundances. \citet{kingsburgh94} also used this equation, with $\alpha$~$=$~3.

A different method of determining the ICF for sulphur was calculated by~\citet{kwitter01}, who considered newer atomic data and incorporated the charge exchange rates into the ICF values. They used the equation 

\begin{equation}
\textrm{ICF(S}^{+} + \textrm{S}^{2+}) = \textrm{exp}[-0.017 + 0.18\beta - 0.11\beta^{2} + 0.072\beta^{3}]
\label{eq:khs12}
\end{equation}

\noindent where $\beta$~$=$~log(O/O$^+$).

The models of \citet{delgadoinglada14a} calculate ICFs for the S/O ratio, and multiply by the O/H abundance:

\begin{equation}
\textrm{log ICF((S}^{+} + \textrm{S}^{2+})/\textrm{O}^{+}) = \frac{-0.02 - 0.03\omega - 2.31\omega^{2} + 2.19\omega^{3}}{0.69 + 2.09\omega - 2.69\omega^{2}}
\label{eq:diso12}
,\end{equation}

\begin{equation}
\textrm{log ICF(O}^{+} + \textrm{O}^{2+}\textrm{)} = \frac{0.08\nu + 0.006\nu^{2}}{0.34 - 0.27\nu}
\label{eq:odi12}
\end{equation}

\noindent where $\nu$ is as defined in Equation (\ref{eq:dine24}) and $\omega$ $=$ O$^{2+}$/(O$^+$~+~O$^{2+}$). In all cases, $\omega$~$>$~0.5 and for K3-67, M2-2 and Y-C~2-5, $\omega$~$>$~0.95.

The sulphur abundances calculated using these ICFs are shown in Table~\ref{tab:SICFs}. Again, the values calculated using empirical ICFs compare well with those from photoionisation models. ICFs obtained with Equation~(\ref{eq:kbs12}) are often large when $\alpha$~$=$~2, and provide much greater estimates than those of the compared studies. However, the agreement is greatly improved when $\alpha$~$=$~3. The uncertainties in the ICF calculated from~\citet{kwitter01} are larger due to their propagation, but the ICFs themselves are smaller, with most of them having values of $\leqslant$~1.35. The only exception to this is K3-90, which has a stronger radiation field than the others (S$^{3+}$/S$^{2+}$~$=$~5.4). Despite these uncertainties, the S/H values calculated from Equation~(\ref{eq:khs12}) show a good agreement with the other values. The abundances calculated using ICFs from \citet{delgadoinglada14a} are similar in that their uncertainties are relatively high, though in almost all cases they show better agreement with the abundances calculated using empirically determined ICFs than those from \citet{kwitter01}.

\begin{table*}
\centering
\caption{Comparison of the sulphur abundances and ICFs used in this study with those in which ICFs from other sources have been applied (see Sect.~\ref{sec:s432}).}
\label{tab:SICFs}
\begin{tabular}{ c c c c c c c c c c c }
\hline\hline
 & \multicolumn{2}{c}{This work} & \multicolumn{4}{c}{B80 / KB94} & \multicolumn{2}{c}{KH01} & \multicolumn{2}{c}{DI14}  \\
 & S/H & & \multicolumn{2}{c}{$\alpha$~=~3} & \multicolumn{2}{c}{$\alpha$~=~2} & S/H &  & S/H & \\
PN & $\times 10^{-6}$ & ICF(S) & S/H $\times 10^{-6}$ & ICF(S) & S/H $\times 10^{-6}$ & ICF(S) & $\times 10^{-6}$ & ICF(S) & $\times 10^{-6}$ & ICF(S) \\
\hline 
K3-66 & 1.7~$\pm$~0.6$^{* \dagger \ddagger}$ & 1.18 & 1.4~$\pm$~1.0$^{* \dagger}$ & 1.10 & 1.6~$\pm$~1.2$^{* \dagger}$ & 1.29 & $^{\#}$ & ... & $^{\#}$ & ... \\
K3-67 & 2.2~$\pm$~0.7$^{* \dagger \ddagger}$ & 1.18 & 2.1~$\pm$~0.9$^{* \dagger}$ & 1.98 & 3.6~$\pm$~1.5$^{* \dagger}$ & 3.38 & 1.2~$\pm$~0.7$^{* \dagger}$ & 1.22 & 1.9~$\pm$~1.3$^{* \dagger}$ & 1.91 \\
K3-70 & 3.7~$\pm$~1.8$^{* \dagger \ddagger}$ & 1.18 & 3.3~$\pm$~0.6$^{* \dagger}$ & 1.33 & 4.5~$\pm$~1.0$^{* \dagger}$ & 1.79 & 2.7~$\pm$~1.1$^{* \dagger}$ & 1.09 & 3.3~$\pm$~2.0$^{* \dagger}$ & 1.31 \\
K3-90 & 2.7~$\pm$~1.2$^{* \dagger \ddagger}$ & 1.18 & 3.6~$\pm$~1.8$^{* \dagger}$ & 5.39 & 10.1~$\pm$~5.7$^{* \dagger}$ & 15.3 & 1.9~$\pm$~1.7$^{* \dagger}$ & 2.86 & 12.4~$\pm$~10.2$^{* \dagger}$ & 18.7 \\
K4-48 & 2.2~$\pm$~0.7$^{* \dagger \ddagger}$ & 1.18 & 2.3~$\pm$~0.8$^{* \dagger}$ & 1.51 & 3.3~$\pm$~1.3$^{* \dagger}$ & 2.20 & 1.7~$\pm$~0.9$^{* \dagger}$ & 1.13 & 2.3~$\pm$~1.5$^{* \dagger}$ & 1.52 \\
M1-6 & 1.6~$\pm$~0.6$^{\dagger}$ & 1.11 & $^{\#}$ & ... & $^{\#}$ & ... & $^{\#}$ & ... &$^{\#}$ & ... \\
M1-7 & 4.4~$\pm$~2.6$^{* \dagger \ddagger}$ & 1.18 & 3.6~$\pm$~0.9$^{* \dagger}$ & 1.21 & 4.6~$\pm$~1.3$^{* \dagger}$ & 1.53 & 3.2~$\pm$~1.4$^{* \dagger}$ & 1.07 & 3.4~$\pm$~1.9$^{* \dagger}$ & 1.15 \\
M1-9 & 2.0~$\pm$~1.0$^{* \dagger \ddagger}$ & 1.00 & 2.2~$\pm$~1.5$^{* \dagger}$ & 1.21 & 2.9~$\pm$~2.0$^{* \dagger}$ & 1.53 & $^{\#}$ & ...  & $^{\#}$ & ...  \\
M1-14 & 2.5~$\pm$~1.0$^{* \dagger \ddagger}$ & 1.00 & 2.7~$\pm$~1.5$^{* \dagger}$ & 1.14 & 3.3~$\pm$~1.9$^{* \dagger}$ & 1.39 & 2.6~$\pm$~2.1$^{* \dagger}$ & 1.06 & 2.6~$\pm$~2.4$^{* \dagger}$ & 1.04 \\
M1-16 & 1.8~$\pm$~0.9$^{* \dagger \ddagger}$ & 1.18 & 1.7~$\pm$~0.5$^{* \dagger}$ & 1.37 & 2.3~$\pm$~0.8$^{* \dagger}$ & 1.90 & 1.3~$\pm$~0.6$^{* \dagger}$ & 1.10 & 1.7~$\pm$~1.0$^{* \dagger}$ & 1.38 \\
M2-2 & 2.1~$\pm$~0.5$^{* \dagger \ddagger}$ & 1.18 & 1.3~$\pm$~0.2$^{* \dagger}$ & 2.47 & 2.5~$\pm$~0.5$^{* \dagger}$ & 4.72 & 0.72~$\pm$~0.25$^{* \dagger}$ & 1.35 & 1.2~$\pm$~0.7$^{* \dagger}$ & 2.28 \\
\hline
\end{tabular}
\tablefoot{B80~$=$~\citet{barker80}; KB94~$=$~\citet{kingsburgh94}; KH01~$=$~\citet{kwitter01}; DI14~$=$~\citet{delgadoinglada14a}. In all cases, ICFs are applied to our sulphur ionic abundances from Table~\ref{tab:SAbuns}. Superscript symbols show the ions considered in the calculations: $^{*} =$~S$^+$; $^{\dagger} =$~S$^{2+}$; $^{\ddagger} =$~S$^{3+}$. $^{\#}$ Uncertainties $>$100\%, likely due to uncertainties in the ICF and S$^+$ abundances given from~\citet{henry10}. }
\end{table*}

\subsubsection{Argon}    \label{sec:ar433}
In optical spectra, the Ar$^{2+}$, Ar$^{3+}$ and Ar$^{4+}$ ions can all be observed. The most abundant of these ions is thought to be Ar$^{2+}$, though this largely depends on the radiation field of the source.

Argon abundances calculated by~\citet{kingsburgh94} applied the following ICFs.

\begin{equation}
\textrm{ICF(Ar}^{2+} + \textrm{Ar}^{3+} + \textrm{Ar}^{4+}\textrm{)} = \frac{1}{1 - \textrm{(N}^{+}/\textrm{N)}}
\label{eq:arkb234}
,\end{equation}

\begin{equation}
\textrm{ICF(Ar}^{2+}\textrm{)} = 1.87~\pm~0.41
\label{eq:arkb2}
.\end{equation}

\citet{kwitter01} built upon Equation (\ref{eq:arkb234}) by considering the ICF when only the Ar$^{2+}$ and Ar$^{3+}$ ionic states could be observed:

\begin{equation}
\textrm{ICF(Ar}^{2+} + \textrm{Ar}^{3+}\textrm{)} = \frac{1}{1 - \textrm{(N}^{+}/\textrm{N)}} \times \frac{\textrm{He}^{+} + \textrm{He}^{2+}}{\textrm{He}^{+}}
\label{eq:arkh23}
.\end{equation}

\citet{delgadoinglada14a} calculated their abundances in terms of Ar/O, before multiplying by the O/H abundance calculated with the ICF from Equation~(\ref{eq:odi12}):

\begin{equation}
\textrm{log ICF}\bigg(\frac{\textrm{Ar}^{2+}}{\textrm{O}^{+} + \textrm{O}^{2+}}\bigg) = \frac{0.03\omega}{0.4 - 0.3\omega} - 0.05 \hskip 2em (0.5 < \omega < 0.95)
\label{eq:ardi2b}
.\end{equation}

\begin{table*}
\centering
\caption{Comparison of the argon abundances and ICFs used in this study with those in which ICFs from other sources have been applied (see Sect.~\ref{sec:ar433}).}
\label{tab:ArICFs}
\begin{tabular}{ c c c c c c c c c }
\hline\hline
 & \multicolumn{2}{c}{This work} & \multicolumn{2}{c}{KB94} & \multicolumn{2}{c}{KH01} & \multicolumn{2}{c}{DI14} \\
 & Ar/H & & Ar/H & & Ar/H & & Ar/H & \\
PN & $\times 10^{-7}$ & ICF(Ar) & $\times 10^{-7}$ & ICF(Ar) & $\times 10^{-7}$ & ICF(Ar) & $\times 10^{-7}$ & ICF(Ar) \\
\hline 
K3-66 & 11.2~$\pm$~4.4$^{* \dagger}$ & 1.35 & 10.5~$\pm$~2.5$^{\dagger}$ & 1.87 & ... & ... & 6.2~$\pm$~5.0$^{\dagger}$ & 1.10 \\
K3-67 & 6.7~$\pm$~2.0$^{* \dagger \ddagger}$ & 1.00 & 8.4~$\pm$~2.0$^{\dagger}$ & 1.87 & 6.5~$\pm$~3.3$^{\dagger \ddagger}$ & 1.05 & 7.2~$\pm$~5.6$^{\dagger}$ & 1.60 \\
K3-70 & 13.0~$\pm$~6.6$^{* \dagger \ddagger}$ & 1.00 & 15.3~$\pm$~4.1$^{\dagger}$ & 1.87 & 16.7~$\pm$~9.5$^{\dagger \ddagger}$ & 1.47 & 11.5~$\pm$~10.0$^{\dagger}$ & 1.40 \\
K3-90 & 7.5~$\pm$~3.4$^{* \dagger \ddagger \#}$ & 1.00 & 4.2~$\pm$~1.1$^{\dagger}$ & 1.87 & ... & ... & ... & ... \\
K4-48 & 16.2~$\pm$~5.3$^{* \dagger \ddagger}$ & 1.00 & 19.1~$\pm$~4.6$^{\dagger}$ & 1.87 & 18.8~$\pm$~8.5$^{\dagger \ddagger}$ & 1.30 & 15.3~$\pm$~12.3$^{\dagger}$ & 1.50 \\
M1-6 & 18.3~$\pm$~7.2$^{* \dagger}$ & 1.35 & 9.7~$\pm$~2.3$^{\dagger}$ & 1.87 & ... & ... & ... & ... \\
M1-7 & 16.2~$\pm$~9.5$^{* \dagger}$ & 1.00 & 19.1~$\pm$~4.6$^{\dagger}$ & 1.87 & 20.3~$\pm$~7.2$^{\dagger \ddagger}$ & 1.51 & 12.9~$\pm$~9.6$^{\dagger}$ & 1.26 \\
M1-9 & 7.5~$\pm$~3.7$^{* \dagger \ddagger}$ & 1.00 & 9.1~$\pm$~2.4$^{\dagger}$ & 1.87 & 6.7~$\pm$~4.0$^{\dagger \ddagger}$ & 1.31 & 5.9~$\pm$~4.8$^{\dagger}$ & 1.21 \\
M1-14 & 12.3~$\pm$~4.9$^{* \dagger}$ & 1.35 & 14.3~$\pm$~3.5$^{\dagger}$ & 1.87 & 11.0~$\pm$~8.6$^{\dagger \ddagger}$ & 1.44 & 8.8~$\pm$~8.3$^{\dagger}$ & 1.14 \\
M1-16 & 22.2~$\pm$~11.1$^{* \dagger \ddagger \#}$ & 1.00 & 21.0~$\pm$~6.1$^{\dagger \ddagger \#}$ & 1.18 & 25.4~$\pm$~11.7$^{\dagger \ddagger}$ & 1.46 & 20.2~$\pm$~16.2$^{\dagger}$ & 1.45 \\
M2-2 & 8.7~$\pm$~2.0$^{* \dagger \ddagger}$ & 1.35 & 6.2~$\pm$~1.5$^{\dagger}$ & 1.87 & 7.0~$\pm$~4.6$^{\dagger \ddagger}$ & 1.10 & 5.8~$\pm$~4.9$^{\dagger}$ & 1.74 \\
Y-C 2-5 & 5.4~$\pm$~3.4$^{\dagger}$ & 1.64 & 6.2~$\pm$~1.6$^{\dagger}$ & 1.87 & ... & ... & 11.3~$\pm$~8.8$^{\dagger}$ & 3.42 \\
\hline
\end{tabular}
\tablefoot{KB94~$=$~\citet{kingsburgh94}; KH01~$=$~\citet{kwitter01}; DI14~$=$~\citet{delgadoinglada14a}. In all cases, ICFs are applied to our argon ionic abundances from Table~\ref{tab:ArAbuns}. Superscript symbols show the ions considered in the calculations: $^{*} =$~Ar$^+$; $^{\dagger} =$~Ar$^{2+}$; $^{\ddagger} =$~Ar$^{3+}$; $^{\#} =$~Ar$^{4+}$.}
\end{table*}

Table~\ref{tab:ArICFs} shows a comparison between the Ar/H abundances using ICFs from all studies. In general, the empirical ICFs agree well with those from each of the three comparative studies. The uncertainties in the abundances calculated from the ICFs of \citet{delgadoinglada14a} are large, particularly for PNe with strong ionisation fields ($\omega$~$\gtrsim$~0.95). This is likely due to the ICF only requiring the ionic abundance of Ar$^{2+}$. 

Overall, the abundances calculated with empirically determined ICFs in all three considered elements compare well with those using ICFs from the literature. The main advantage of using the empirical method is that there is no need for large ICFs due to the number of ionisation states for which we have data. While much larger ICFs are applied in the literature to account for missing ionisation states that contribute more to the elemental abundances, the resulting values relate well to those calculated empirically in which some or all of these missing ionisation states have been observed.

\subsection{Galactocentric distances}
Distances to Galactic PNe are known to be notoriously difficult to measure owing to the variation of bolometric luminosity and effective temperature values within the sources. We considered distances relative to the Galactic centre and exclusively observed PNe towards the anti-centre, which somewhat reduces the errors relative to those of their respective heliocentric distances ($R_h$).

We converted from $R_h$ to $R_g$ using the following equation: 

\begin{multline}
|\overrightarrow{R_{g}}| = ([R_{h} \cos(b) \cos(l) - R_{\odot}]^{2} + R_{h}^{2} \cos^{2}(b) \sin^{2}(l) \\ + R_{h}^2 \sin^{2}(b))^{0.5}
\label{eq:dist}
\end{multline} 

\noindent where we take $R_{\odot}$~$=$~8.0~$\pm$~0.5~kpc.

Table~\ref{tab:23PNe} shows the distance values used for each PN. In this paper, we primarily adopt values from~\citet{frew16} as measured through statistical means, though where possible we prioritised the use of directly determined distances from~\citet{giammanco11}. Table~\ref{tab:rhcomp} shows the $R_h$ values of the PNe in the anti-centre sample with distance values given by~\citet{giammanco11}, alongside those from~\citet{frew16}. While there are strong disagreements in most cases, the abundances towards the anti-centre remain lower than elsewhere in the Galactic disk. The choice of data set will not affect the overall conclusions as each case presents a similar level of dispersion around the overall gradient, part of which comes from uncertainties in the distance measurements.

\begin{table}
\centering
\caption{Comparison of heliocentric distances from~\citet{giammanco11} and~\citet{frew16}. }
\label{tab:rhcomp}
\begin{tabular}{ c c c }
\hline\hline
 & $R_h$ / kpc & $R_h$ / kpc \\
PN & (Giammanco+'11)  & (Frew+'16) \\
\hline
K3-65 & 3.7 & 13.0 \\
K3-68 & 2.2 & 7.4 \\
K3-69 & $>$6.0 & ... \\
K3-70 & $>$6.0 & 15.8 \\
K3-71 & 2.5 & 18.2 \\
K3-90 & $<$1.0 & 7.0 \\
M1-6 & 2.0 & 5.2 \\
M2-2 & $>$2.0 & 5.2 \\
\hline
\end{tabular}
\end{table}

The uncertainties quoted by~\citet{frew16} are estimated to be $\sim$~20--30\%, which seem small for statistical values, though larger uncertainties will not affect the outcomes of our discussion.

\section{Discussion}  \label{sec:Discussion}

\subsection{Comparison of abundances with literature}
In Tables~\ref{tab:NeAbuns}--\ref{tab:ArAbuns}, we have compared the elemental abundances of neon, sulphur, and argon with those available from literature. We accounted for average uncertainties of 17.5\% in neon abundances for the missing Ne$^{3+}$ ionisation state,  25\% in sulphur abundances for S$^{+}$ and S$^{4+}$, and 30\% in argon for Ar$^{3+}$ and occasionally Ar$^{+}$. We used the sample of \citet{henry10} as our main comparison, as this is a recent study involving 12 of our 23 PNe, though we also considered the abundances shown within the study of~\citet{sterling08}. These all involved the use of optical spectra. 

In most cases, there is good agreement between all sets of abundances. Where we have lower abundance values than in literature, this likely comes from over-estimated ICFs from optical studies; for example, the neon abundance of K3-90 from literature (Table~\ref{tab:NeAbuns}) has an ICF of 23.8 (from Equation \ref{eq:kbne2}) applied to an uncorrected value relatively similar to ours~\citep{henry10}, resulting in our elemental abundances disagreeing with theirs by a factor of eight. The neon ICF was high because the oxygen ICF was comparatively large (22.3), which was calculated from the ratio of (He$^{+}$~+~He$^{2+}$)/He$^+$ abundances~\citep{kwitter01}. \citet{kingsburgh94} give the same oxygen ICF to the power of 2/3, which would have decreased the ICF from 22.3 to 7.9 and the neon abundance to (4.7~$\pm$~2.6)~$\times$~10$^{-5}$. This is narrowly within the uncertainty range of our Ne/H value ((1.8~$\pm$~0.8)~$\times$~10$^{-5}$).

Conversely, our neon abundance of M4-18 is a factor of 20 greater than that given in literature by~\citet{demarco99}. Their abundance was calculated using the 12.8~$\mu$m [Ne~{\tiny II}] line flux from~\citet{aitken82} (3.6~$\times$~10$^{-12}$~erg~cm$^{-2}$~s$^{-1}$), which compares well with our  value, 3.3~$\times$~10$^{-12}$~erg~cm$^{-2}$~s$^{-1}$. However, the value of log~$I({H\beta})$ used by~\citet{demarco99} is lower than ours ($-$11.44 and $-$11.24, respectively), accounting for a discrepancy of factor 1.6. We have used $I(H\beta)$ and $C(H\beta)$ values from~\citet{henry10}, but we were unable to directly compare our neon abundances for this source as these authors were unable to observe the dominant Ne$^+$ ion, and Ne$^{2+}$ was barely observable, if at all (they gave its ionic abundance with a 90\% uncertainty). 

The source M1-1 also shows relatively large discrepancies between our abundance and those given in literature for neon and argon. By considering its high ionisation field, from which we can observe Ne$^{5+}$, we expect that Ne$^{3+}$ and Ar$^{3+}$ will contribute significantly to their overall abundances. \citet{aller86} estimate the abundances of Ne$^{2+}$, Ne$^{3+}$ and Ne$^{4+}$ in M1-1, obtaining values of a factor of approximately three lower than our values, with Ne$^{+3}$ contributing 40\% of the total of those three ions (taken from measurements of the 2424$\AA$ [Ne~{\tiny IV}] line). They also estimate Ar$^{3+}$ to contribute 31\% of the total argon abundance. ICFs of 1.12 and 2.1, for neon and argon respectively, are applied to their overall abundances. Even so, there are still significant discrepancies between our sets of abundance values.
The abundances of the anti-centre PNe agree well with those from literature. For the few cases in which there are discrepancies between these values, they are likely to come from the larger ICFs used within the literature.

\subsection{The abundance gradient}  \label{sec:abungrad}
\begin{figure*}
\centering
\includegraphics[width=0.33\textwidth]{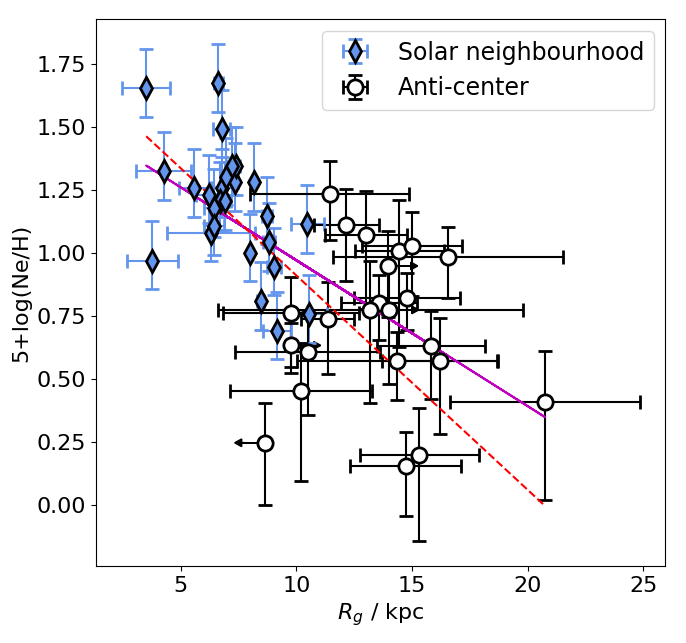}\hfill
\includegraphics[width=0.33\textwidth]{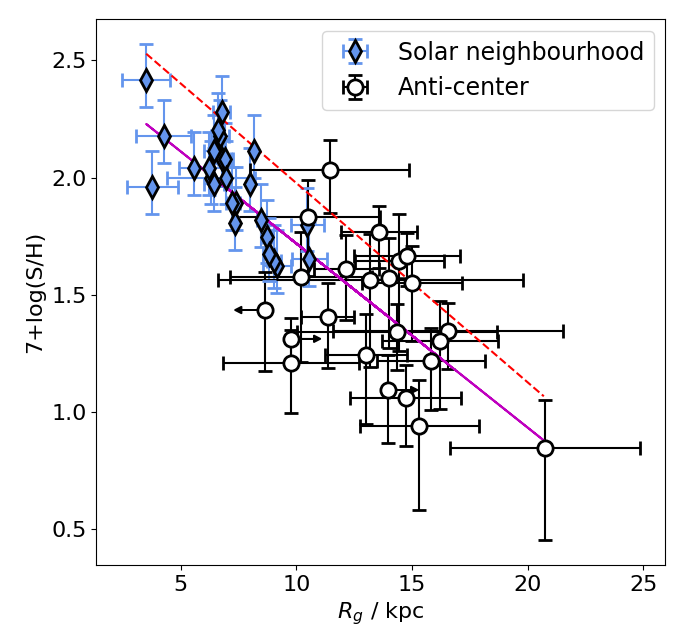}\hfill
\includegraphics[width=0.33\textwidth]{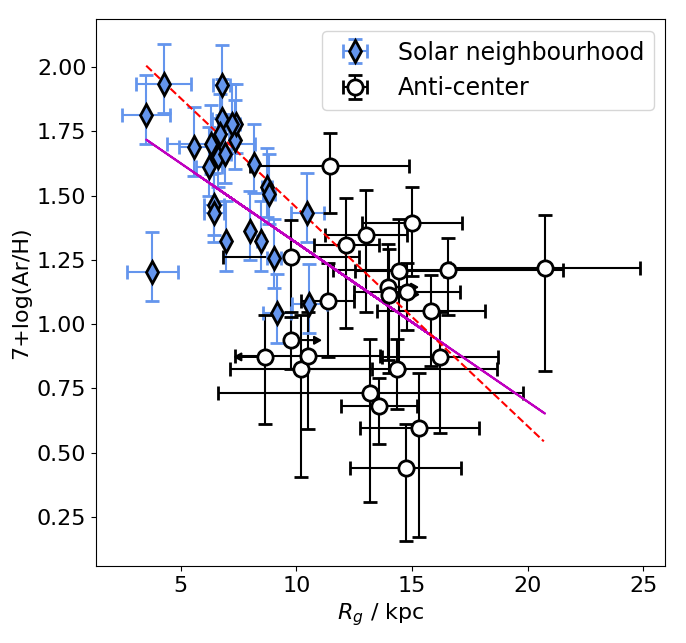}
\caption{Abundance gradients of neon, sulphur, and argon in the Milky Way. The dashed lines represent the oxygen abundance gradient from within the Galactic disk with a slope of $-$0.085~dex/kpc~\citep{pottasch06}, passing through the solar value at 8.0~kpc~\citep{asplund05}. The solid lines represent the line of best fit in each plot, with gradients of  $-$0.058~$\pm$~0.021, $-$0.079~$\pm$~0.012 and $-$0.062~$\pm$~0.023~dex/kpc respectively.}
\label{fig:abungrads}
\end{figure*}

Figure~\ref{fig:abungrads} shows the abundances of neon, sulphur, and argon plotted against $R_g$ for the PNe in the anti-centre sample and for those of~\citet{pottasch06} from the solar neighbourhood. Both of these samples were analysed and reduced in a similar way.

\begin{figure*}
\centering
\includegraphics[width=0.33\textwidth]{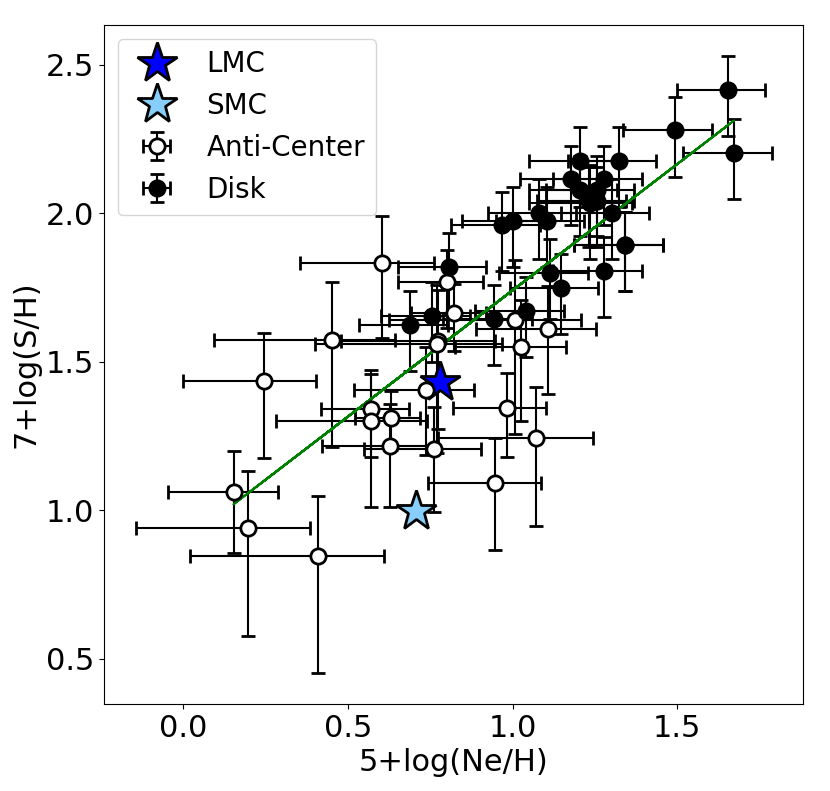}\hfill
\includegraphics[width=0.33\textwidth]{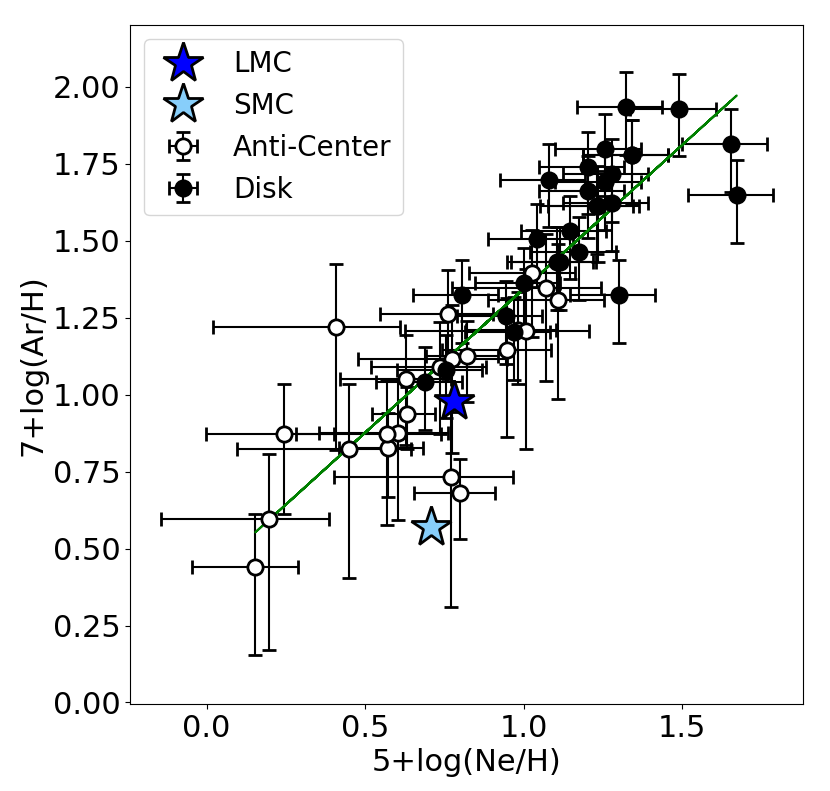}\hfill
\includegraphics[width=0.33\textwidth]{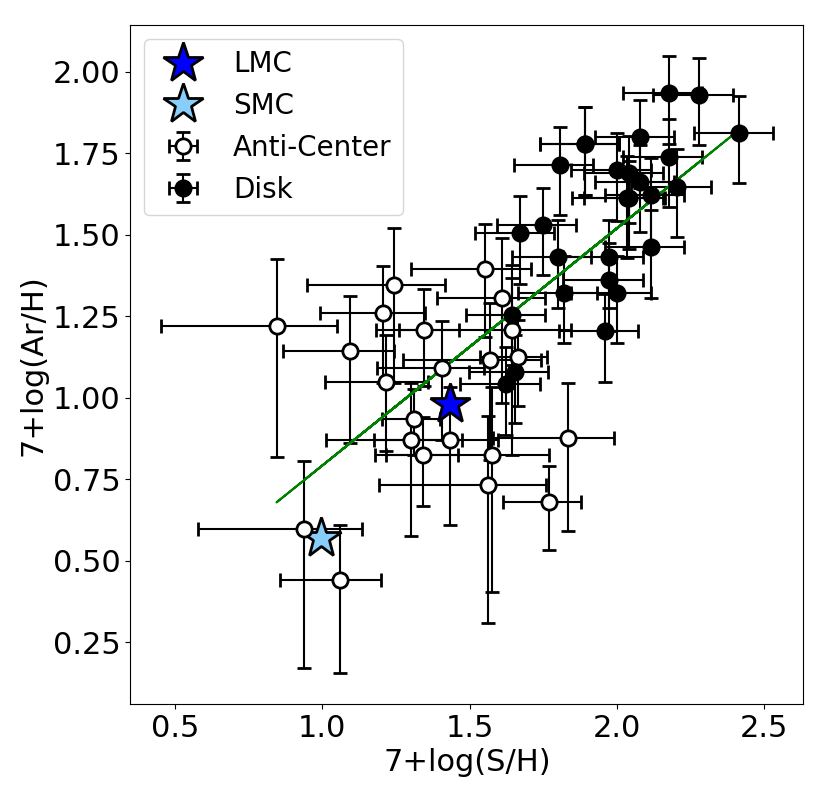}
\caption{Plots comparing the neon, sulphur, and argon abundances. The Galactic Disk sample was analysed by~\citet{pottasch06}. The average abundances for the LMC and SMC are also shown, with those of neon and sulphur obtained from~\citet{bernardsalas08} and argon from~\citet{leisy06}.}
\label{fig:abuncomps}
\end{figure*}

The abundances are lower than those of the Galactic bulge and the solar neighbourhood, and are consistent with a continuation of the metallicity gradient up to $R_g$~$\sim$~20~kpc, albeit with a large dispersion within the data. However, when analysing the neon and argon anti-centre data separately from the solar neighbourhood data, there is no clear correlation, with Pearson correlation coefficients ($R_{PCC}$) of $\sim$~$-0.05$ in each case. Hence, we cannot discern with our data whether there is a gradient in the anti-centre ($R_g$~$>$~10~kpc). For sulphur, the anti-centre data show $R_{PCC}$~$=$~$-0.45$ with a corresponding p-value of 0.032, showing that there is a slight negative correlation in the anti-centre that is statistically significant. Together with the solar neighbourhood data, $R_{PCC}$~$=$~$-0.64$, $-0.82$ and $-0.66$ for neon, sulphur, and argon respectively.

Table~\ref{tab:gradcomp} compares the radial metallicity gradients from a selection of studies over multiple wavebands, including PNe, H~{\tiny II} regions, young B-type stars and Cepheid variables. Our analysis includes studies of neon, sulphur, and argon gradients. We also considered oxygen; even though its abundance changes over the course of stellar evolution in PNe, \citet{pottasch06} suggested that the oxygen gradient of sources that had not undergone hot bottom burning was identical to those of the other three elements from their PN sample. Our metallicity gradient slopes compare well with those from most other studies, though ours have greater uncertainties which arise from the dispersion at greater $R_g$, primarily due to the uncertainties in the distance measurements. Despite this, the slopes calculated from the PN studies of \citet{maciel99}, \citet{pottasch06} and this work are typically steeper than those of other source types. Our metallicity gradients are consistent with a continuation at high $R_g$, though the slopes suggest an eventual flattening or steepening with distance is possible, particularly for neon and argon. While some of the studies we considered rule out flattening as a possibility, the studies of~\citet{andrievsky02b,andrievsky02a,andrievsky02c} and \citet{luck03}, analysing Cepheid variables, find discontinuities in the abundance gradient with $R_g$. They show that the gradient is seen to be steeper in regions closer to the Galactic centre ($R_g$~$\sim$~4--6~kpc) and towards the anti-centre ($R_g$~$\sim$~10--15~kpc) for 25 different elements, including oxygen and sulphur.
Several studies that analyse sources beyond $\sim$~15~kpc (e.g.~\citealt{rolleston00,lemasle13,fernandezmartin17}; this work) show steeper gradients than those found in most other studies. While this appears to agree with the findings of \citet{andrievsky02b,andrievsky02a,andrievsky02c}, there are relatively few sources at these distances in each named study, hence the effect of the large distance sources on the respective gradients is likely to be minimal. By factoring in the large uncertainties in distance measurements for most sources in these samples, we do not find that the abundance gradients steepen with $R_g$ in the direction of the anti-centre.

Analysis of the time evolution of the Galactic abundance gradient from this PN sample is also difficult, due to the large uncertainties in abundances and Galactocentric distances. Our data are consistent with a continuation of the gradient at large distances, so there is no suggestion that the inner and outer disks of the Milky Way evolved separately (e.g.~\citealt{stanghellini10,kubryk15b}).

\begin{table*}
\centering
\caption{Comparison of several abundance gradient studies which use various sources.}
\label{tab:gradcomp}
\begin{tabular}{ c c c c c c }
\hline\hline
Study & Sources & Waveband & Element & Slope (dex/kpc) & $R_g$ range (kpc) \\
\hline
\multirow{3}{*}{This work} & \multirow{3}{*}{PNe} & \multirow{3}{*}{IR, Optical} & Ne & $-$0.058~$\pm$~0.021 & 3--21* \\
 & & & S & $-$0.079~$\pm$~0.012 & 3--21* \\
 & & & Ar & $-$0.062~$\pm$~0.023 & 3--21* \\ \vspace{0.15mm} \\
\multirow{4}{*}{\citet{maciel99}} & \multirow{4}{*}{PNe} & \multirow{4}{*}{IR, Optical} & O & $-$0.058~$\pm$~0.007 & 4--14 \\
 & & & Ne & $-$0.036~$\pm$~0.010 & 4--14 \\
 & & & S & $-$0.077~$\pm$~0.011 & 4--13 \\
 & & & Ar & $-$0.051~$\pm$~0.010 & 4--13 \\ \vspace{0.15mm} \\
\citet{pottasch06} & PNe & IR, Optical, UV & O, Ne, S, Ar & $-$0.085 & 3--11 \\ \vspace{0.15mm} \\
\citet{maciel15} & PNe & multiple & O & $-$0.025~$\pm$~0.006 & 0--15 \\ \vspace{0.15mm} \\
\multirow{2}{*}{\citet{martinhernandez02}} & \multirow{2}{*}{H~{\tiny II} regions} & \multirow{2}{*}{Optical} & Ne & $-$0.039~$\pm$~0.007 & 0--14 \\
 & & & Ar & $-$0.045~$\pm$~0.011 & 0--13 \\ \vspace{0.15mm} \\
\citet{esteban17} & H~{\tiny II} regions & Optical & O & $-$0.040~$\pm$~0.005 & 5--17$^{\dagger}$ \\ \vspace{0.15mm} \\
\multirow{3}{*}{\citet{fernandezmartin17}} & \multirow{3}{*}{H~{\tiny II} regions} & \multirow{3}{*}{Optical} & O & $-$0.053~$\pm$~0.009 & 11--18 \\
 & & & S & $-$0.106~$\pm$~0.006 & 11--18 \\
 & & & Ar & $-$0.074~$\pm$~0.006 & 11--18 \\ \vspace{0.15mm} \\
\citet{fitzsimmons92} & B-type stars & multiple & O & $-$0.03~$\pm$~0.02 & 5--14 \\ \vspace{0.15mm} \\
\citet{rolleston00} & B-type stars & Optical & O & $-$0.067~$\pm$~0.008 & 6--18 \\ \vspace{0.15mm} \\
\multirow{2}{*}{\citet{andrievsky02a}} & \multirow{2}{*}{Cepheids} & \multirow{2}{*}{IR} & O & $-$0.022~$\pm$~0.009 & 6--11 \\
 & & & S & $-$0.051~$\pm$~0.008 & 6--11 \\ \vspace{0.15mm} \\
\citet{lemasle13} & Cepheids & Optical, near-IR & S & $-$0.095~$\pm$~0.015 & 4--19$^{\ddagger}$ \\ \vspace{0.15mm} \\
\multirow{4}{*}{\citet{henry04}} & \multirow{4}{*}{\parbox{3cm}{\centering PNe, H~{\tiny II} regions, B-type stars, Cepheids}} & \multirow{4}{*}{multiple} & O & $-$0.037~$\pm$~0.008 & 0--18 \\
 & & & Ne & $-$0.044~$\pm$~0.014 & 2--14 \\
 & & & S & $-$0.048~$\pm$~0.010 & 0--17 \\
 & & & Ar & $-$0.030~$\pm$~0.010 & 2--17 \\
\hline
\end{tabular}
\tablefoot{Studies with `multiple' wavebands use data from several references. * Includes data from~\citet{pottasch06}; $^{\dagger}$ Includes data from~\citet{esteban15}; $^{\ddagger}$ Includes data from~\citet{luck11}.}
\end{table*}

\subsection{$\alpha$-process elements}
In the evolution of low- to intermediate-mass stars, the abundances of elements heavier than carbon are generally not affected, except for those formed during the slow neutron-capture process (known as the $s$-process) which can occur during the AGB phase (e.g.~\citealt{lugaro12}). As a result, the abundances of neon, sulphur, and argon should trace each other. Figure~\ref{fig:abuncomps} shows that there is good agreement between these abundances, hence proving that they do trace each other well, though it is clear that the plot of sulphur against argon shows a greater dispersion. This could be explained by the need to account for two ions for argon (Ar$^{+}$ and Ar$^{3+}$), for which Ar$^{3+}$ can be the dominant ion.

Included on these plots are the abundances of Ne, S and Ar from the Magellanic Clouds. The mean abundance values for neon and sulphur were taken from an IR \emph{Spitzer} sample of PNe from~\citet{bernardsalas08}, and those for argon were taken from an optical PN sample from~\citet{leisy06}. The anti-centre sample shows abundances scattered around the LMC metallicity for each of the three elements, with few reaching values below those of the SMC.

The sulphur anomaly is the term coined by~\citet{henry04}, used to describe the observed sulphur abundances in PNe being lower than those of H~{\tiny II} regions at the same metallicity (see also e.g.~\citealt{henry12}). It was originally suggested that this could be explained by the lack of measured emission lines of ionisation states of S$^{3+}$ and above in optical spectra and the need to account for them, particularly as S$^{3+}$ can be a key stage of ionisation for sulphur. The sulphur anomaly has been seen in multiple galaxies; \citet{garciarojas16} observe this anomaly from four H~{\tiny II} regions with abundances greater than most of the thirteen PNe in their sample from NGC 6822. \citet{shaw12} find the anomaly in the Magellanic Clouds from a combination of IR, optical and UV data, and \citet{shingles13} find the anomaly in the Milky Way from the PN sample of~\citet{pottasch10}, also from spectra in the same wavebands, compared to the ISM trend of H~{\tiny II} regions and blue compact galaxies from the optical sample of~\citet{milingo10}. 

\begin{figure}
\centering
\includegraphics[width=\columnwidth]{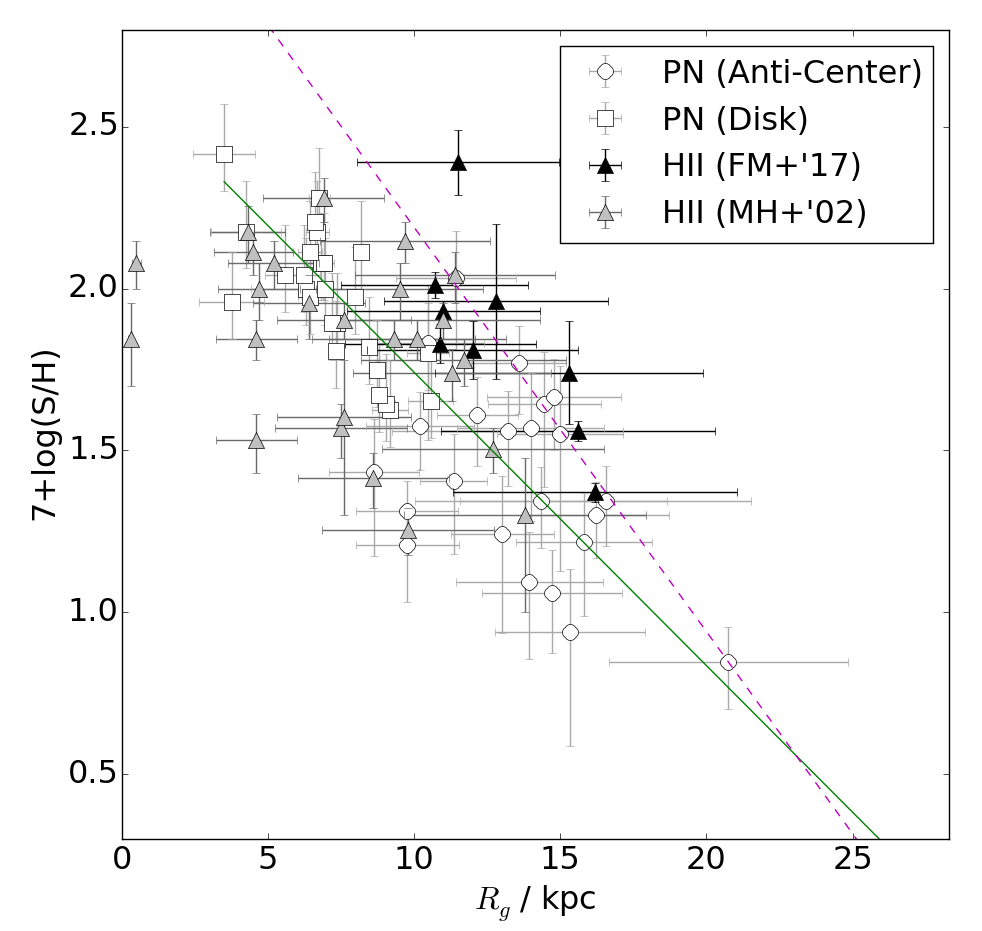}
\caption{Sulphur abundances of our PNe alongside samples of H~{\tiny II} regions from~\citet{fernandezmartin17} (FM+'17) and \citet{martinhernandez02} (MH+'02), and solar neighbourhood PNe from~\citet{pottasch06}. The solid line is the line of best fit for PN abundances. The dashed line represents that for the H~{\tiny II} regions of the FM+'17 sample.}
\label{fig:hii}
\end{figure}

In Fig.~\ref{fig:hii}, we compare the sulphur abundances of PNe in the Galactic anti-centre and solar neighbourhood~\citep{pottasch06} with the sulphur abundances in two samples of Galactic H~{\tiny II} regions, one derived from IR data~\citep{martinhernandez02} and the other from optical data~\citep{fernandezmartin17}. Both of these samples cover a similar range of Galactocentric distance values to the anti-centre PNe.

The IR H~{\tiny II} region data from~\citet{martinhernandez02} agree well with the PN abundances, inferring that the sulphur anomaly is not observed from these data. However, their H~{\tiny II} region abundances disagree with the interstellar and solar values of sulphur by a factor of approximately two to four. Based on this, they argue that their abundances are underestimated by up to a factor of four, which they ascribe to uncertainties in their $n_e$ and $T_e$ values, with the lack of S$^+$ abundance values from their IR data from \emph{ISO} accounting for a further $\sim$~15\% discrepancy. We note that since the release of the paper of~\citet{martinhernandez02}, the most widely used solar sulphur abundance value from the literature shows a decrease of $\sim$~20\% from the value they used \citep{snow96,asplund09}, though both the abundances of the anti-centre PNe and the H~{\tiny II} regions remain low in comparison.

Comparing the infrared PN abundances with the optical H~{\tiny II} region abundance data of~\citet{fernandezmartin17}, with Galactocentric distances of 11--17~kpc, shows a clear discrepancy between the two sets of data. The PN abundances are lower than those of the H~{\tiny II} regions by a factor of approximately two, as shown by the two lines of best fit in Fig.~\ref{fig:hii}. In this case, we clearly observe the sulphur anomaly.

\section{Summary}  \label{sec:Summary}
We have presented an infrared spectroscopic study of 23 PNe in the Galactic anti-centre with $R_{g}$ values of 8--21~kpc using \emph{Spitzer} IRS to determine the abundances of neon, sulphur, and argon in a region that is a priori assumed to be metal-poor.

We have calculated the abundances in two ways: using empirically calculated ICFs from a combination of IR and optical data, and using more well-established ICFs from the literature. We find that the two methods produce similar results; the empirical ICFs consider a wider range of ionic states and are therefore small in value. We find that the abundances of neon, sulphur, and argon are lower in the anti-centre than those in the solar neighbourhood. The metallicity gradients of these elements seem to continue beyond $R_g$~$=$~10~kpc despite a large spread of data values. The abundances of the $\alpha$-process elements trace each other well, though there is a slightly larger dispersion between those of sulphur and argon.

\emph{Spitzer} IRS has enabled the study of abundances from observations of PNe in the bulge, disk and halo of the Milky Way, as well as in nearby galaxies (primarily the Magellanic Clouds) at infrared wavelengths. With its greater sensitivity, the James Webb Space Telescope (\emph{JWST}) will be able to continue to obtain spectra for PNe as far as the Local Group of galaxies, enabling us to carry out abundance studies over a wider range of parameter space. In addition, \emph{JWST} will be able to spatially resolve PNe in the Milky Way, allowing us to investigate how the gas and dust content varies within these nebulae.

\begin{acknowledgements}
We thank the anonymous referee for their feedback and comments. This research has made use of the NASA/IPAC Infrared Science Archive, which is operated by the Jet Propulsion Laboratory, California Institute of Technology, under contract with the National Aeronautics and Space Administration. This work has been supported by a Marie Curie FP7 CIG grant under project number 630861 (FEASTFUL).
\end{acknowledgements}

\bibliographystyle{aa} 
\bibliography{Antigc_AbundanceB1_lang}

\end{document}